\documentclass[pra,superscriptaddress,twocolumn,superscriptaddress,amsmath]{revtex4}

 \usepackage{graphicx}
\usepackage{float}
\usepackage{epstopdf,cancel}
\usepackage{epsf,latexsym,bbm,euscript}
\usepackage{amssymb,amsmath}
\usepackage{mathtools} 
\usepackage{times,graphics}
\usepackage{soul,xcolor}
\usepackage{epsfig}
\usepackage{verbatim}
\usepackage{todonotes}

\usepackage[normalem]{ulem}

\def\6{\langle}
\def\9{\rangle}

\newcommand\co{{\cal{O}}}

\newcommand\hn{{\hat{n}}}

\newcommand\vv{{\vec{v}}}
\newcommand\vx{{\vec{x}}}
\newcommand\va{{\vec{a}}}
\newcommand\vd{{\vec{d}}}

\def\vbe{{\vec{\beta}}}

\def\sg{\textsl{g}}

\newcommand\ma{{\mathfrak{a}}}
\newcommand\md{{\mathfrak{d}}}
\newcommand\mrin{{\mathrm{in}}}

\newcommand\vch{{\vec{\chi}}}
 \newcommand\vnu{{\vec{\nu}}}

\def\half{\tfrac{1}{2}}

\newcommand{\pad}{\partial}
\newcommand{\be}{\begin{equation}}
\newcommand{\ee}{\end{equation}}
\newcommand{\ba}{\begin{eqnarray}}
\newcommand{\ea}{\end{eqnarray}}

\newcommand{\defeq}{\vcentcolon=}
\newcommand{\eqdef}{=\vcentcolon}

\usepackage{lipsum}

 \usepackage{url,hyperref}
\hypersetup{colorlinks,linkcolor={blue!55!black},citecolor={red!50!black},urlcolor={blue!45!black}}

\begin{document}

\title{Proposal for an optical interferometric measurement \\ of the gravitational red-shift with satellite systems}
\author{Daniel~R.~Terno}
\email{daniel.terno@mq.edu.au}
\affiliation{School of Mathematical and Physical Sciences, Macquarie University, Sydney NSW 2109, Australia}
\author{Francesco~Vedovato}
\affiliation{Dipartimento di Ingegneria dell'Informazione, Universit\`{a} degli Studi di Padova, Padova 35131, Italy}
\affiliation{Istituto Nazionale di Fisica Nucleare (INFN) --- Sezione di Padova, Italy}
\author{Matteo~Schiavon}
\affiliation{Dipartimento di Ingegneria dell'Informazione, Universit\`{a} degli Studi di Padova, Padova 35131, Italy}
\affiliation{Istituto Nazionale di Fisica Nucleare (INFN) --- Sezione di Padova, Italy}
\affiliation{Sorbonne Universit\'{e}, CNRS, LIP6, F-75005 Paris, France}
\author{Alexander~R.~H.~Smith}
\affiliation{Department of Physics, Saint Anselm College, Manchester, New Hampshire 03102, USA}
\affiliation{Department of Physics and Astronomy, Dartmouth College, Hanover, New Hampshire 03755, USA}
\author{Piergiovanni~Magnani}
\affiliation{Department of Physics, Politecnico di Milano, Milano 20133, Italy}
\author{Giuseppe~Vallone}
\affiliation{Dipartimento di Ingegneria dell'Informazione, Universit\`{a} degli Studi di Padova, Padova 35131, Italy}
\affiliation{Istituto Nazionale di Fisica Nucleare (INFN) --- Sezione di Padova, Italy}
\affiliation{Dipartimento di Fisica e Astronomia, Universit\`a di Padova, via Marzolo 8, 35131 Padova, Italy}
\author{Paolo~Villoresi}
\email{paolo.villoresi@dei.unipd.it}
\affiliation{Dipartimento di Ingegneria dell'Informazione, Universit\`{a} degli Studi di Padova, Padova 35131, Italy}
\affiliation{Istituto Nazionale di Fisica Nucleare (INFN) --- Sezione di Padova, Italy}

\begin{abstract}
The Einstein Equivalence Principle (EEP) underpins all metric theories of gravity. One of its key aspects is the local position invariance (LPI) of non-gravitational experiments, which is captured by the gravitational red-shift. The iconic gravitational red-shift experiment places two fermionic systems, used as clocks, in different gravitational potentials and compares them using the electromagnetic field. However, the electromagnetic field itself can be used as a clock, by comparing the phases acquired by two optical pulses propagating through different gravitational potentials.
A fundamental point in the implementation of a satellite large-distance optical interferometric experiment is the suppression of the first-order Doppler effect, which dominates the weak gravitational signal necessary to test the EEP. Here, we propose a novel scheme to suppress it, by subtracting the phase-shifts measured in the one-way and in the two-way configuration between a ground station and a satellite. We present a detailed analysis of this technique within the post-Newtonian framework and perform some simulations of its performance using realistic satellite orbits and the state-of-the-art fiber technology at the telecom wavelength of 1550~nm.
\end{abstract}

\maketitle

\section{Introduction}
Light, apart from the \textit{ad hoc} applications of corpuscular analogies, is insensitive to the Newtonian gravity. The situation is conceptually very different in general relativity (GR): indeed, in all metric theories of gravity the electromagnetic (EM) wave propagation depends on the spacetime background  \cite{mtw,will:93,wp:book}. In the short wavelength limit light rays \cite{opt-bw}, which are characteristic curves of the wave equation, model classical and quantum beams, as well as trajectories of single photons \cite{opt-bw,q-opt,t:14}.  On curved backgrounds the short-wave asymptotic expansion identifies rays as null geodesics \cite{mtw,wp:book}. However,   near the surface of our planet the gravitational effects appear at the  $c^{-2}$ post-Newtonian order, where $c$ is the speed of light. These second-order terms are often masked by much stronger kinematic effects.

For example, the optical version of the Colella-Overhauser-Werner (COW) experiment~\cite{cole75prl} was proposed in~\cite{zych11nco}. Using communications between a spacecraft and a ground station to realize the Mach-Zehnder interferometer (the experiment was suggested in~\cite{sat:12} as a possible component of the QEYSSAT mission~\cite{QEYSSAT_spie}), it is possible to obtain a quite large gravitationally-induced phase shift,
\begin{equation}
   { \varphi_{\rm gr} =  \Delta U \omega_0 \tau_l  \approx - \frac{gh}{c^2} \frac{2\pi }{\lambda} nl } \ .  \label{phigr}
\end{equation}
In this scheme a photon time-bin superposition~\cite{brendel1999} is sent from a ground station on Earth to a spacecraft. Both terminals are equipped with  a fiber-based interferometer of  equal temporal  imbalance $\tau_l = nl/c$ (with $n=1.5$ the refractive index of the fiber and $l$ the length of the delay line), in order to temporally recombine the two time-bins and obtain an interference pattern depending on the gravitational phase-shift {$\varphi_{\rm gr} = \Delta\omega \tau_l$~\cite{phase-freq}, where the frequency shift $\Delta\omega$ is derived below, in Eq.~\eqref{alphadef}. Here we approximated the difference of the gravitational potential as $g h$, with $g$ the Earth's gravity and $h$ the satellite altitude, and $\lambda = 2\pi c/\omega_0$ is the sent wavelength. The order of magnitude of the gravitational red-shift is about 1~rad, supposing $\lambda = 1550$~nm, $l=1.2$~km and an altitude $h = 1500$~km (which corresponds to $\Delta U \approx -1.3 \cdot 10^{-10}$).
The expected signal lies into a measurable regime, and an optical precision of $\delta\varphi_{\rm gr} \approx 10$~$\mu$rad is experimentally achievable provided a number of detected photons $N$ fulfilling $N \gtrsim 1/\delta\varphi_{\rm gr}^2$~\cite{shot-noise}.

However, the careful analysis of the optical COW in~\cite{bggst:15} showed that the first-order Doppler effect is roughly $10^5$ times stronger than the desired signal $\varphi_{\rm gr}$. Moreover, in this setting the kinematic and gravitational effects are ineludibly linked~\cite{phase-freq}. This  first-order Doppler effect was recently measured by exploiting large-distance precision interferometry along space channels~\cite{padova:16}, which represents a resource for performing fundamental tests of quantum mechanics in space, as in~\cite{Vedovato17,zych11nco,zych12cqg,Yin17, Ren17},  for future space-based scientific missions, such as LISA~\cite{lisa-over}, and space-based quantum cryptography~\cite{Liao17,cubesat,Bedington17,Agnesi18, Khan2018, Liao2018}.

A novel proposal for the extraction of the  gravitational contribution to the phase is the subject of the present work. Our goal --- direct observation of the effects of gravity in an optical interferometric experiment --- is part of the efforts to design new tests of the Equivalence Principle. We now review its formulation and connection to the gravitationally-induced phase, and then outline the structure of the following discussion.

The Einstein Equivalence Principle (EEP) is the foundation of all metric theories of gravity, including  general relativity~\cite{mtw,will:93,wp:book,inertia,will:lrr}. EEP comprises three statements. The first --- {\it Weak Equivalence Principle} --- states that the trajectory of a freely falling test body is independent of its internal composition. The other two statements deal with outcomes of non-gravitational experiments performed in freely falling laboratories where self-gravitational effects are negligible. The second statement --- {\it Local Lorentz Invariance} --- asserts that such experiments are independent of the velocity of the laboratory where the experiment takes place. The third statement --- {\it Local Position Invariance} (LPI) --- asserts that ``the outcome of any local non-gravitational experiment is independent of where and when in the universe it is performed''~\cite{will:lrr}. 

Tests of the ``when'' part of the EEP bound the variability of the non-gravitational constants over  cosmological time scales~\cite{dirac:937, uzan:lrr,LPI:21}. The ``where'' part  was expressed in Einstein's analysis~\cite{einstein:911} of what in modern terms is a comparison of two identical frequency standards in two different locations in a static gravitational field. The so-called {\it red-shift} implied by the EEP affects the locally measured frequencies of a spectral line that is emitted at location 1 with  the proper frequency  $\omega_0$ and then detected at location $2$ with $\omega'$. The red-shift can be parametrized as 
 \begin{equation}
{\frac{\Delta \omega}{\omega_0} = (1 + \alpha) \Delta U + \mathcal{O}(c^{-3})} \ , \label{alphadef}
\end{equation}
where $\Delta \omega := \omega^\prime - \omega_0$  and  $\Delta U := U_2 - U_1$, where  $U_i:=-\phi_i/c^2$ has the opposite sign of the Newtonian gravitational potential $\phi_i$ at the emission ($1$) and detection ($2$), while $\alpha \neq 0$ accounts for possible violations of LPI}.
In principle, $\alpha$ may depend on the nature of the clock that is used to measure the red-shift~\cite{will:lrr,will:93,uzan:lrr}. 
For example, the standard model extension (SME) includes all possible Lorentz- and CPT-violating terms preserving the fundamental $SU(3)\times SU(2)\times U(1)$ gauge invariance and power-counting renormalizability~\cite{ck:98}. The SME contains constrained parameters whose different combinations may lead to $\alpha\neq 0$, as well as different couplings of the Standard Model parameters and gravity~\cite{sme,lli,hc:11}.

A typical red-shift experiment  involves a pair of clocks, naturally occurring~\cite{sunline} or specially-designed~\cite{pr:60,gpa:80,aces,app:18,Delva2018,Herrmann2018,bothwell:22}, whose readings are communicated by EM radiation. 
It should be noted that the leading term in Eq.~\eqref{alphadef} is the same in all metric theories of gravity.  Evaluating   $\Delta\omega/\omega_0$ to a higher order in the post-Newtonian approximation leads to the expressions that depend on the specifics of the theory and are different between general relativity and alternative metric theories of gravity~\cite{will:93,wp:book}. 
Therefore, as we detail in Sec.~\ref{PPN-1}, the absolute violation of LPI in terms of a single parameter is meaningfully defined in the near-Earth experiments only up to the level of $10^{-5}$.

This level of precision of the measurements of $\alpha$ is already well-established \cite{will:lrr,will:93,app:18,Delva2018}. Moreover, comparison of co-located ultra-precise clocks, using two different atoms (hydrogen and cesium) for their working transitions, allowed for a  bound  on the difference $\alpha_\mathrm{H}-\alpha_\mathrm{Cs}$ with the precision of $2\times10^{-7}$~\cite{app:18}.

These estimations of $\alpha$ are based on implicit or explicit assumptions on the standard propagation of the EM radiation~\cite{hc:11}. Furthermore, parameters of the models with dark matter directly coupling to the EM field are also constrained using atomic measurements~\cite{til:15}. As a matter of principle, once the possibility of LPI violation is entertained, there is no reason for it to be the same  for all fields of the Standard Model, and the distinct coefficients in the symmetry-violating terms in SME are generally considered~\cite{ck:98,sme}. Hence, different types of experiments, which employ a single EM-source and compare optical phase-differences between beams of light traversing different paths in a gravitational field, provide a complementary test of LPI.
Our analysis is purely classical. However, it can be adopted to describe the state transformation of photonic qubits.

The rest of this paper is organized as follows. The frequency shift of Eq.~\eqref{alphadef} underpins the phase difference whose extraction   we outline in Sec.~\ref{s2}. This protocol forms a novel test of the EEP exploiting a single EM-source and a double large-distance interferometric measurement performed at two different gravitational potentials. Sec.~\ref{PPN-1} discusses in detail how by comparing the phase-shifts obtained at a satellite and on Earth, it is possible to overcome the first-order Doppler effect and obtain the gravitational contribution. Sec.~\ref{s4} presents simulations that are based on the orbits of existing and proposed satellites, and discusses the current technological limitations of the scheme.

\section{Description of the proposal and of the Doppler-cancellation scheme}\label{s2}

\begin{figure}[!t]
\centering
\includegraphics[width=0.45\textwidth]{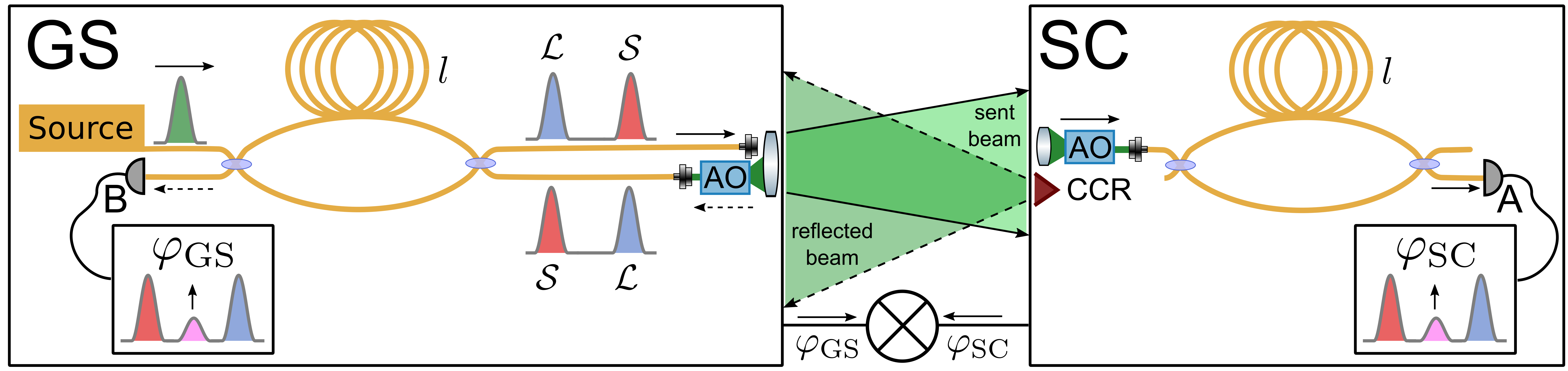} \\
\includegraphics[width=0.45\textwidth]{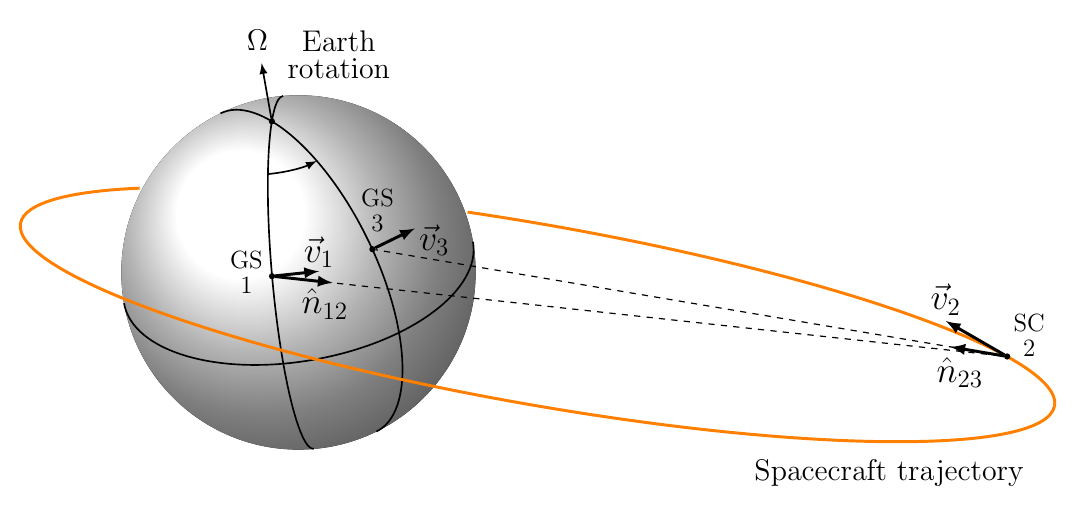} \\
\caption{{\it (top)} Scheme of the proposal. Both the ground station (GS) and the spacecraft (SC) are equipped with a Mach Zehnder interferometer (MZI) of equal delay line $l$ and an adaptive optics (AO) system for fiber injection. {\it (bottom)} Geometry of the experiment: $\vv_1$ is the velocity of the GS at the emission at potential $U_1$; $\vv_2$ is the velocity of the SC at the detection on the satellite at potential $U_2$; $\vv_3$ is the velocity of the GS at the detection of the beam retroreflected by the corner-cube retroreflector (CCR) on the SC, which occurs at potential $U_3 = U_1$. Approximating Earth's angular velocity $\Omega$ as constant, $|\vv_1|^2 = |\vv_3|^2$. Vectors $\hn_{12}$ and $\hn_{23}$ are the Newtonian propagation directions of the light pulses.} 
\label{figure-scheme}
\end{figure}

A possible setup for our proposal is sketched in Fig.~\ref{figure-scheme} and is based on the satellite interferometry experiment realized in~\cite{padova:16}.
Such an interferometric measurement is obtained by sending a light pulse through a cascade of two fiber-based Mach Zehnder interferometers (MZIs) of equal temporal imbalance {$\tau_l$.} After the first MZI the pulse is split into two temporal modes, called \emph{short} ($\mathcal{S}$) and \emph{long} ($\mathcal{L}$) depending on the path taken in the first MZI. The equal imbalance of the two MZIs guarantees that the two pulses are recombined at the output of the second MZI, where they are detected.  The combination of the possible paths the pulses may take leads to a characteristic detection pattern comprising three possible arrival times for each pulse. The first (third) peak corresponds to the pulses that took the $\mathcal{S}$ ($\mathcal{L}$) path in both the MZIs, while the mid peak is due to the pulse that took the $\mathcal{S}$ path in the first interferometer and the $\mathcal{L}$ path in the subsequent one, or viceversa. Hence, interference is expected only in the central peak, due to the indistinguishability of the two possibilities. 

Such an interference is modulated by the phase-difference $\varphi$ accrued in the propagation by the two interfering paths, that depends on the relative motion between the ground station (GS) and the spacecraft (SC), as depicted in Fig.~\ref{figure-scheme}, and on the difference in gravitational potentials, as we will detail in the following. From the ratio of the intensity of the central peak to the lateral ones an estimation of $\varphi$ can be obtained~\cite{padova:16}. 
To realize this interferometric measurement, the coherence time of the source $\tau_c$ must be, at the same time, much shorter than the temporal imbalance $\tau_l \approx \tau_l^{\rm GS} \approx \tau_l^{\rm SC}$ introduced by the single delay line, and longer than the mismatch $\Delta\tau_l := \tau_l^{\rm SC} - \tau_l^{\rm GS}$ between the two interferometers (which cannot be perfectly identical), i.e.
\begin{equation}
    \Delta\tau_l < \tau_c \ll \tau_l \ . \label{inequality_source}
\end{equation}
We will show in Appendix~\ref{appendix_details} how the setting of the source can be chosen such that Eq.~\eqref{inequality_source} is satisfied.

Furthermore, we assumed that a free-space to single-mode fiber coupling system is implemented to guarantee the spatial overlap of the interfering beams and thus resulting in a high visibility (the interferometric visibility is further discussed in Appendix~\ref{appendix_visibility}).
The latter assumption seems to be very demanding from an experimental point of view. However, it was recently demonstrated that it is possible to couple into single-mode fibers a laser beam coming from satellites~\cite{Takenaka2012,Wright2015}. Indeed, by using an adaptive optics (AO) system~\cite{Wright2015}, it is possible to correct the wavefront distortion induced by turbulence and to mitigate losses and intensity fluctuations at the receiver. We note that, as discussed below, the phase-difference $\varphi$ is not affected by turbulence. 
More technical details on the experimental setup, attesting the feasibility of our proposal within a decade, are given in Appendix~\ref{appendix_details}.

The Doppler-cancellation scheme is based on the fact that the one-way phase-difference $\varphi_{\rm SC}$ contains both the first-order Doppler and higher-order terms including the gravitational contribution $U_2 - U_1 \equiv U_{\rm SC} - U_{\rm GS}$, while the two-way one, $\varphi_{\rm GS}$, contains only Doppler terms, since the gravitational contribution is cancelled out at the leading order in the two-way trip. The first-order Doppler terms are eliminated by manipulating the corresponding data sets from the GS and SC in a manner similar to the time-delay interferometry techniques in Ref.~\cite{tdi}. The key feature allowing for this is that the ratio of first-order Doppler terms in $\varphi_{\rm SC}$ and $\varphi_{\rm GS}$ is exactly equal to two (see below).

Hence, using the linear combination
\begin{equation}
    S := \varphi_{\rm SC} - \half  \varphi_{\rm GS} \label{eq_signal}
\end{equation} 
 of the two phase-differences $\varphi_{\rm SC}$ and $\varphi_{\rm GS}$, that are obtained from an interferometric measurement of the kind described above, a bound on $\alpha$ will be retrieved. It  parallels the data processing in the Gravity Probe A experiment~\cite{gpa}. Here $\varphi_{\rm SC}$ is measured at detector A located on the SC, while $\varphi_{\rm GS}$ at detector B located at the GS, by exploiting the reflection of the sent beam obtained with a corner-cube retroreflector (CCR) mounted on the SC~(Fig.~\ref{figure-scheme}).

 The explicit form of the signal is derived in the next Section and in Appendix~\ref{appendix_calculation}, resulting in 
\begin{align}
\frac{S}{\omega_0 \tau_l} &= (1+\alpha)(U_2 - U_1) +\half(\beta_2^2- \beta_1^2) \nonumber \\ &\quad -\vbe_1\cdot(\vbe_1-\vbe_2) 
- (\md_2^2-\md_1^2)- T (\hn_{12} \cdot \va_1) \nonumber \\ &\quad\quad - \left((\vbe_2-\vbe_1)^2-(\md_2-\md_1)^2\right) \frac{\tau_l}{4T}\ , \label{FinalSignal_explicit}
\end{align}
 where $\alpha$ parametrizes the violation of LPI, $\vbe_i:= \vv_i/c$, $\md_i := \hn_{12} \cdot \vbe_i$, $T$ is the zeroth order time-of-flight between the GS and the SC, $\va_1$ is the centripetal acceleration of the GS at 1, and the other quantities are specified in Fig.~\ref{figure-scheme}.

\begin{figure*}[t]
    \centering
        \includegraphics[width = 0.32\textwidth]{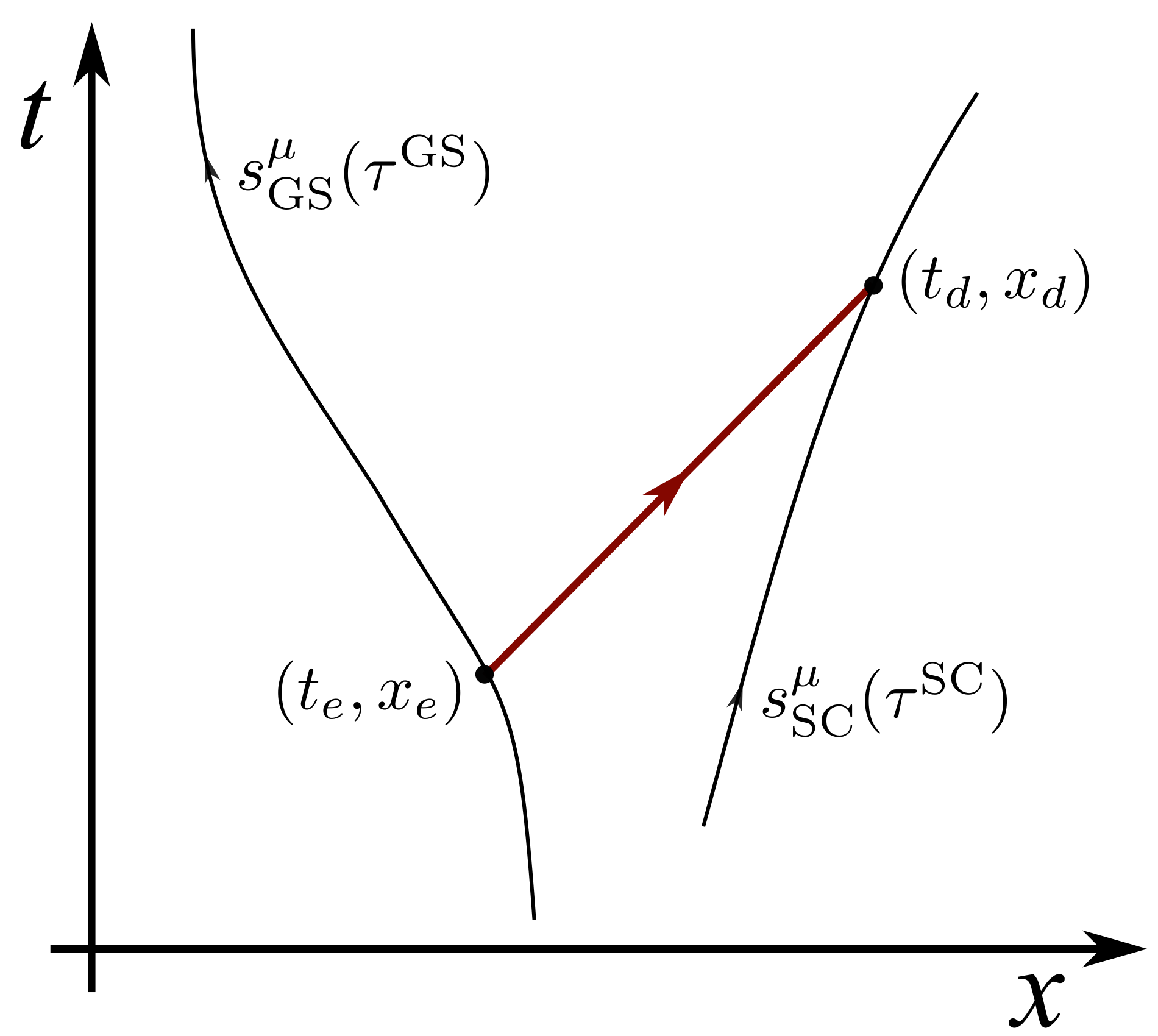} 
    \includegraphics[width = 0.32\textwidth]{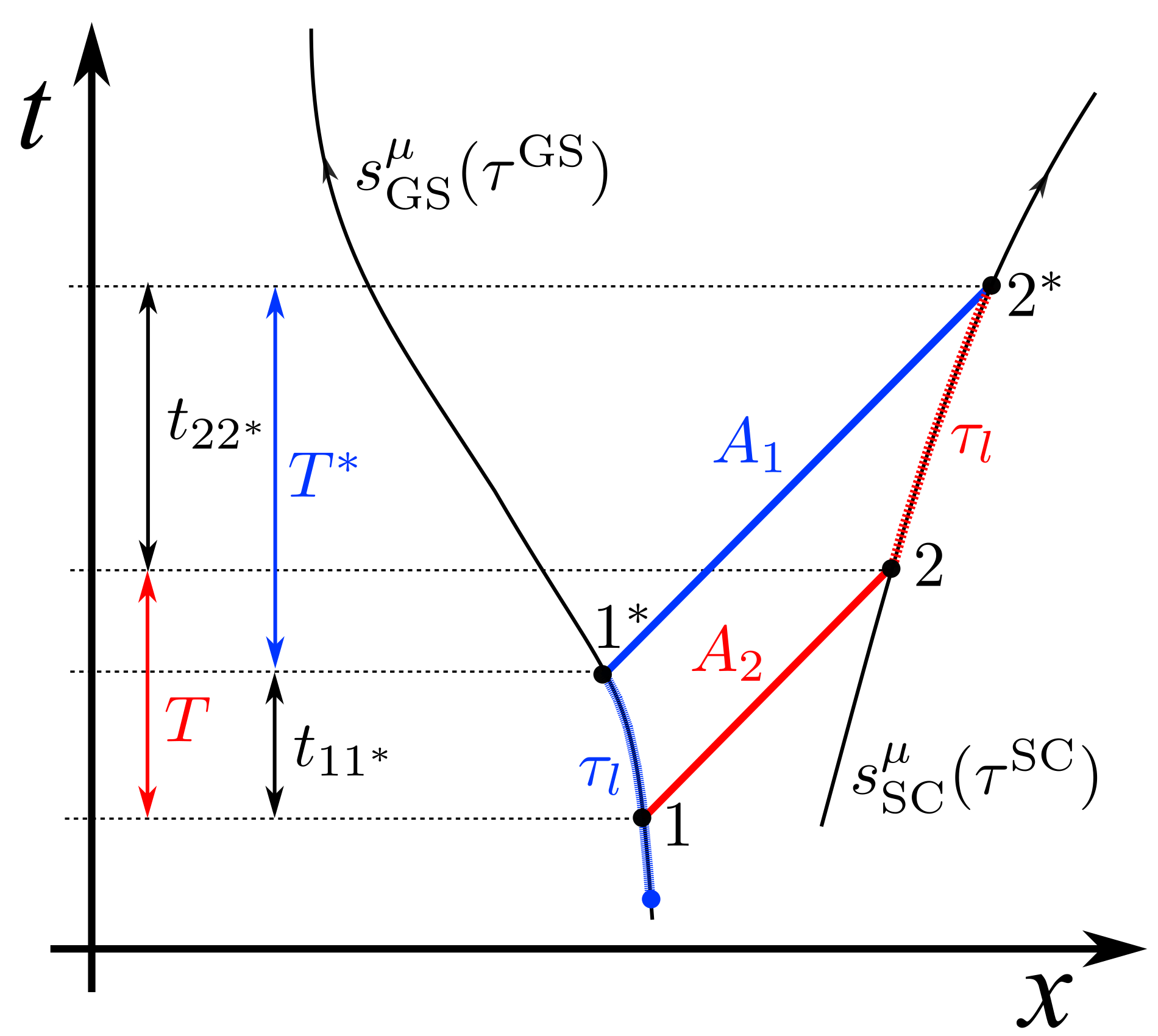} 
    \includegraphics[width = 0.32\textwidth]{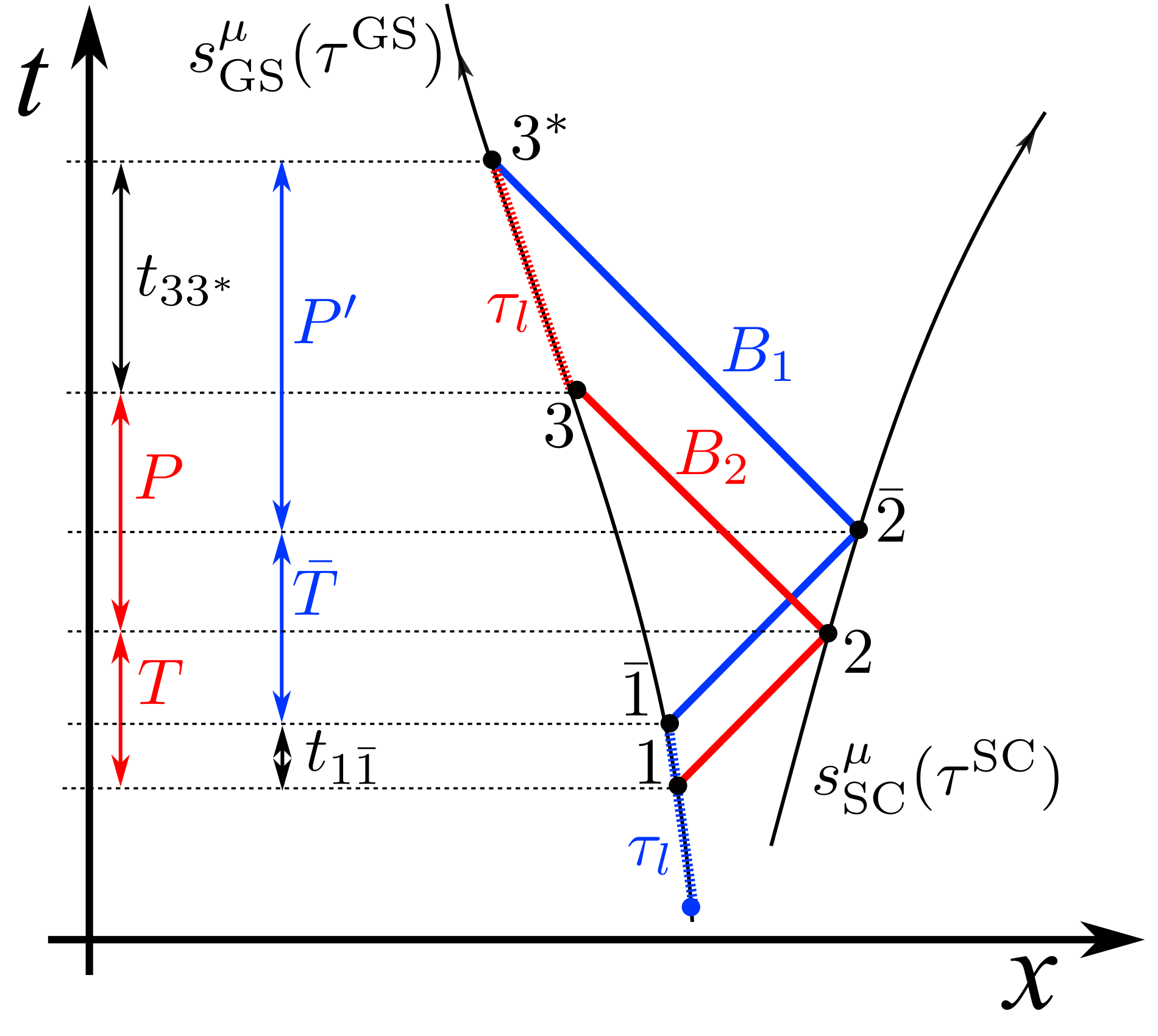}
    \caption{{\it (left)} Spacetime diagram with a single null geodesic segment connecting the emission and detection points lying along the  {ground station and the spacecraft}  worldlines ($s_{\rm GS}^{\mu}$ and $s_{\rm SC}^{\mu}$, respectively). {\it (center)} Spacetime diagram for the one-way phase-shift. We define: $T := t_2 - t_1$ (zeroth order flight-of-time from the GS to the SC) and $T^* := t_{2^*} - t_{1^*}$. {\it (right)} Spacetime diagram for the two-way phase-shift. We define: $\bar{T} := t_{\bar{2}} - t_{\bar{1}}$, $P' := t_{3^*} - t_{\bar{2}}$, and $P := t_3 - t_2$. The coordinates $(t,x)$  refer to the GRF. The $x$-axis represents all three spatial directions.}
    \label{fig:mink}
\end{figure*}

\section{Phase-shift estimation in the PPN approximation}\label{PPN-1}

{\it Notation.---}We present the detailed analysis of the phases to be measured by exploiting the Parametrized Post Newtonian (PPN) formalism~\cite{will:93,wp:book,mtw} using the notation of~\cite{bggst:15}. The order of expansion is labelled by the parameter $\epsilon$, which is taken equal to 1 at the end of calculation.
The PPN formalism applied to near-Earth experiments implies $\epsilon \approx 10^{-5}$, since Earth's gravitational potential is defined to be of the order $\epsilon^2$ and $U_{\oplus}= G M_\oplus/(c^2 R_\oplus) \approx 10^{-10}$~\cite{will:93} (the subscript $\oplus$ refers to Earth).  It is worth noticing that the absolute value of the GS and SC velocities $v_i/c$ are also bounded by $10^{-5}$, thus being of the first order in $\epsilon$. Moreover, another scale-parameter is important in our problem, and it is given by the ratio $\mu := \tau_l/T$ between the delay-line imbalance and the zeroth-order time-of-flight. For an imbalance of {$l=1.2$~km}, as used in the following, we have {$\mu \approx 10^{-3}$}.

At this level of precision, we can ignore the effects of the gravitational field of other bodies in the Solar System, approximate the spacetime around the Earth as static, and consider only the leading (i.e. second order in $\epsilon$) post-Newtonian effects. Thus, the non-vanishing components of the metric in the PPN approximation are~\cite{will:93,wp:book,mtw,inertia}
\begin{equation}
\sg_{00}=-1+  2 U \ ,  \quad \sg_{ij}=\delta_{ij}\big(1+  2 U\big) \ , 
\end{equation}
where the gravitational potential around Earth includes the quadrupole term~\cite{inertia}
\begin{equation}
U :=U(r, \theta) =  \frac{GM_{\oplus}}{c^2 r}\left(1-\half J_2\frac{R_{\oplus}^2}{r^2}(3\cos^2\theta-1)\right)
\end{equation}
with $J_2=1.083\times 10^{-3}$ the normalized quadrupole moment.  
The off-diagonal terms in the PPN-metric are of the order $\epsilon^3$, while the next-order correction to $\sg_{00}$ is of the order $\epsilon^4$~\cite{will:93,wp:book}. Taking these and higher-order terms into account allows to obtain the frequency-shift with an arbitrary precision. Unlike the universal  $\epsilon^2$ term, the $\epsilon^3$ and higher-order terms  depend on the specific EEP-conforming metric theory used \cite{will:93,wp:book}.

Unit (Euclidean) vectors $\hn_{ij}$ describing light propagation direction carry double subscripts indicating the starting ($i$) and ending ($j$) points of the geodesic segment followed by the pulse. More details on light propagation in the PPN formalism are reported in Appendix~\ref{section_lightprop}.

Since we deal with short time intervals, we use an Earth-centered inertial system as the standard reference frame with coordinates $(t, \vx)$. For brevity we refer to this system as the ``global'' reference frame (GRF), distinguishing it from the local frames that are established at the GS and the SC along their wordlines parametrized by the proper times $\tau^{\rm GS}$ and $\tau^{\rm SC}$ [Fig.~\ref{fig:mink}(left)],  which are distinguished by superscripts. Quantities that are expressed in the GRF usually will carry no superscripts. On the other hand, the subscripts refer to the location of a particular event: for example 1 and 3 occur at the GS, while 2 happens at the SC (see Fig.~\ref{figure-scheme}).  In the following calculation we use the coordinates $\vx_1$ and $\vx_2$, the velocities $\vv_1$ and $\vv_2$, and accelerations $\va_1$ and $\va_2$ at the points $1$ and $2$ and suppose the time-of-flight $T$ as known. 

Coordinate-time and proper-time intervals are defined as $t_{ij} := t_{j} - t_i$ and $\tau_{ij}:= \tau_j - \tau_i$ respectively,  and they can be related by using the line element
  \be
  -d\tau^2=(-1+2U)dt^2+\frac{v^2}{c^2}dt^2+\co(\epsilon^3) \ ,
  \ee
where $\tau$ is the proper time of the local observer (at the GS or SC) that moves with the velocity $\vv = c \vbe$. Hence
\be
\tau_{ij}=\left(1-\half \beta_i^2-U_i\right)t_{ij} \ , \label{eq_trasf_frame}
\ee
that is exact at the order $\co(\epsilon^2)$, provided that $v_i t_{ij}\lesssim \epsilon r_i$ and $a_i t_{ij}\lesssim\epsilon v_i$.

{\it How to evaluate the phase-shift.---}The most effective way to carefully estimate the phase-shift for the interfering beams in the scheme proposed in Fig.~\ref{figure-scheme} is to use the spacetime diagrams of Fig.~\ref{fig:mink}~\cite{phase-freq}. By describing light wave propagation using geometric optics~\cite{mtw,opt-bw}, we have that the scalar amplitude of a monochromatic wave can be written as $\psi (t, \vx) = A(t, \vx)e^{i\Phi(t, \vx)}$, where the phase $\Phi(t, \vx)$ is scalar function satisfying the eikonal equation, which amounts to the Hamilton-Jacobi equation for massless particles~\cite{mtw,opt-bw}. If we consider a single null geodesic segment that connects two points belonging to two timelike trajectories --- as $(t_e, \vx_e)$ and $(t_d, \vx_d)$ in Fig.~\ref{fig:mink}(left) --- we have that the accrued phase can be evaluated indifferently at the emission ($e$) or detection ($d$) point:
\begin{equation}
    \phi[e \rightarrow d] = \phi^{\rm GRF}(t_d, \vx_d) = \phi^{\rm GRF}(t_e, \vx_e) \ .
\end{equation}
Since the phase is a scalar, we can evaluate it in either the local frames established at the SC or at the GS according to
\begin{align}
   \phi[e \rightarrow d] &= \phi^{\rm SC}[\tau_d^{\rm SC}, \vx_d^{\rm SC}(\tau_d^{\rm SC})]  \nonumber \\ 
   &= \phi^{\rm GS}[\tau_e^{\rm GS}, \vx_e^{\rm GS}(\tau_e^{\rm GS})] = \phi_0 - \omega_0 \tau_e^{\rm GS} \ .  \label{eq_recipe}
\end{align}
In the above expression we explicited the form of the phase in the GS-frame, where the emitted frequency is  $\omega_0 := - u^\mu_{\rm GS} k_\mu$, with $k_\mu:= -\pad_\mu\Phi$ the 4-wavevector, $u^\mu_{\rm GS}$ the 4-velocity of the frame and $\phi_0$ is some initial phase. 

In our setting, this recipe implies to back-propagate the light trajectory from the final detection point ($2^*$ for the one-way measurement and $3^*$ for the two-way one) to the GS worldline, and  we have also to take into account the presence of the delay line (of proper time $\tau_l$) in the path [see Fig.~\ref{fig:mink}]. It is worth noticing that, since the two waves associated to the two possible paths are required to interfere at the same spacetime event, the back-propagation implies that the two points where the phase is estimated at the GS are actually two different spacetime events for the two paths. Furthermore, we can apply the machinery described above to pulses of light, since they are obtained as superposition of plane waves~\cite{phase-freq}.

{\it One-way phase-difference.} The spacetime diagram of the two beams $A_1$ and $A_2$ interfering after the one-way trip at the point $2^* := (t_{2^*}, \vx_{2^*})$ is represented in Fig.~\ref{fig:mink}(center). 
$A_2$ is the path followed by the pulse that leaves the GS at $1$, reaches the SC at 2 and ends at $2^*$ by taking the delay line  on the satellite just before the detection. Hence, the phase-shift at the point $2^*$ given the path $A_2$, taking into account the delay line ($\rm d.l.$) and the back-propagation ($\rm b.p.$), is 
\begin{align}
    \phi[A_2] &= \phi^{\rm SC}[2^* | A_2] \nonumber \\
    &\overset{\rm d.l.}{=}\phi^{\rm SC}[\tau_{2}^{\rm SC} := \tau_{2^*}^{\rm SC} - \tau_l, \vx^{\rm SC}(\tau_{2}^{\rm SC})] \nonumber\\
 &\overset{\rm b.p.}{=}\phi^{\rm GS}[\tau_{1}^{\rm GS}, \vx^{\rm GS}(\tau_{1}^{\rm GS})] \nonumber \\
 &\overset{\eqref{eq_recipe}}{=} \phi_0-\omega_0 \tau_{1}^{\rm GS}  \ . 
\end{align}
On the other hand, the path $A_1$ is the one followed by the pulse that arrives at $2^*$ while leaving the GS at $1^*$ after having took the delay line on the ground. Thus, its accrued phase is
\begin{align}
    \phi[A_1] &= \phi^{\rm SC} [2^*|A_1] \nonumber\\
    &\overset{\rm b.p.}{=} \phi^{\rm GS}[\tau_{1^*}^{\rm GS} , \vx^{\rm GS}(\tau_{1^*}^{\rm GS})]
    \nonumber 
    \\
    &\overset{\rm d.l.}{=}
    \phi^{\rm GS}[\tau_{1^*}^{\rm GS} - \tau_l, \vx^{\rm GS}(\tau_{1^*}^{\rm GS} - \tau_l)] \nonumber 
    \\
    &\overset{\eqref{eq_recipe}}{=}  \phi_0-\omega_0 (\tau_{1^*}^{\rm GS} - \tau_l) \ .
\end{align}

The phase-difference for the one-way measurement realized the SC is given by 
\begin{align}
    \varphi_{\rm SC} &:= \phi[ A_2] - \phi[ A_1 ] = \omega_0
    (\tau_{1^*}^{\rm GS}
    -\tau_{1}^{\rm GS} -\tau_l) \ ,
\end{align}
where 
$\tau_{1^*}^{\rm GS}-\tau_{1}^{\rm GS}\equiv
\tau_{11^*}^{\rm GS}$ is related
to coordinate time interval $t_{11^*}$
by Eq.~\eqref{eq_trasf_frame} and 
\begin{equation}
    t_{11^*} + T^* = T + t_{22^*} \  \label{eq_time_oneway}
\end{equation}
holds, with $\tau_{22^*}^{\rm SC} \equiv \tau_l$.
In the Appendix~\ref{sec_back_oneway} we \ evaluate $\varphi_{\rm SC}$ by expanding the unknown quantities  in powers of $\epsilon$ (these are the time-of-flight $T^*$ of the delayed pulse, its Newtonian propagation direction $\hn_{1^{\!*}2^{\!*}}$ and the coordinate-time interval $t_{11^*}$) and by using the equations describing the motion of the SC and the light propagation in the PPN approximation. We finally obtain
\begin{equation}
    \varphi_{\rm SC} = -\omega_0 T_1 + \varphi_{\rm SC}^{(2)} \ ,
\end{equation}
where the first-order Doppler is given by
\begin{equation}
T_1 = \hn_{12}\cdot(\vbe_2-\vbe_1)\tau_l \ .
\end{equation}
 The detailed calculation and  the explicit form of the second-order term $\varphi_{\rm SC}^{(2)}$  {are} given in the Appendix~\ref{sec_back_oneway}.

{\it Two-way phase-difference.} The spacetime diagram of the two beams $B_1$ and $B_2$  interfering after the two-way trip  at the space-time event $3^* := (t_{3^*}, \vx_{3^*})$ is represented in Fig.~\ref{fig:mink}(right). Analogously to the one-way shift, for the $B_2$ path (delay-line on the ground just before the detection) we have that 
\begin{align}
    \phi[B_2] &= \phi^{\rm GS}[3^*| B_2] \nonumber \\
    &\overset{\rm d.l.}{=} \phi^{\rm GS}[\tau_{3}^{\rm GS} := \tau_{3^*}^{\rm GS} - \tau_l, \vx^{\rm GS}(\tau_{3}^{\rm GS})] \nonumber \\ 
    &\overset{\rm b.p.}{=} \phi^{\rm GS}[\tau_{1}^{\rm GS}, \vx^{\rm GS}(\tau_{1}^{\rm GS})] \nonumber\\
    &\overset{\eqref{eq_recipe}}{=} \phi_0 -\omega_0 \tau_{1}^{\rm GS}  \ ,
\end{align}
while for the path $B_1$ (delay-line on the ground at the start) we have that 
\begin{align}
    \phi[B_1] &= \phi^{\rm GS}[3^*| B_1] \nonumber \\
    &\overset{\rm b.p.}{=} \phi^{\rm GS}[\tau_{\bar{1}}^{\rm GS}, \vx^{\rm GS}(\tau_{\bar{1}}^{\rm GS})] \nonumber \\
    &\overset{\rm d.l.}{=} \phi^{\rm GS}[ \tau_{\bar{1}}^{\rm GS} - \tau_l, \vx^{\rm GS}(\tau_{\bar{1}}^{\rm GS} - \tau_l)] \nonumber \\ 
    &\overset{\eqref{eq_recipe}}{=} \phi_0 - \omega_0 (\tau_{\bar{1}}^{\rm GS} - \tau_l)  \ .
\end{align} 
Hence, the phase-difference for the two-way measurement realized at the GS is given by 
\begin{equation}
    \varphi_{\rm GS} := \phi[B_2] - \phi[B_1] = \omega_0
    (\tau_{\bar{1}}^{\rm GS} - \tau_1^{\rm GS}
     -\tau_l) \ ,
\end{equation} 
where  $\tau_{\bar{1}}^{\rm GS} - \tau_1^{\rm GS} \equiv \tau_{1\bar{1}}^{\rm GS}$ is related to $t_{1\bar{1}}$ by Eq.~\eqref{eq_trasf_frame} and 
\begin{equation}
t_{1\bar{1}} + \bar{T} + P' = T + P +  t_{33^*} \   \label{eq_time_twoway}
\end{equation}
with $\tau_{33^*}^{\rm GS} \equiv \tau_l$. With a procedure analogous to the one of the one-way phase-shift, we finally obtain 
\begin{equation}
  \varphi_\mathrm{GS}= -2\omega_0 T_1+   \varphi_\mathrm{GS}^{(2)} \ , 
\end{equation} 
where $\varphi_\mathrm{GS}^{(2)}$ and the detailed calculation are explicitly given in the Appendix~\ref{sec_back_twoway}.

The first-order term $\varphi_{\rm GS}^{(1)}:= -2\omega_0 T_1$ is exactly what has been measured in~\cite{padova:16}. As anticipated above, the ratio of the first-order terms in $\varphi_{\rm SC}$ and $\varphi_{\rm GS}$ is exactly two, thus allowing for the Doppler-cancellation strategy that is summarised in Eq.~\eqref{eq_signal}.

The effect of the length mismatch between the loops are the main practical limitation of the scheme and are discussed in Section~\ref{s4} and Appendix~\ref{appendix_dlines}.

\begin{figure*}[!t]
\centering
  \includegraphics[width=\textwidth]{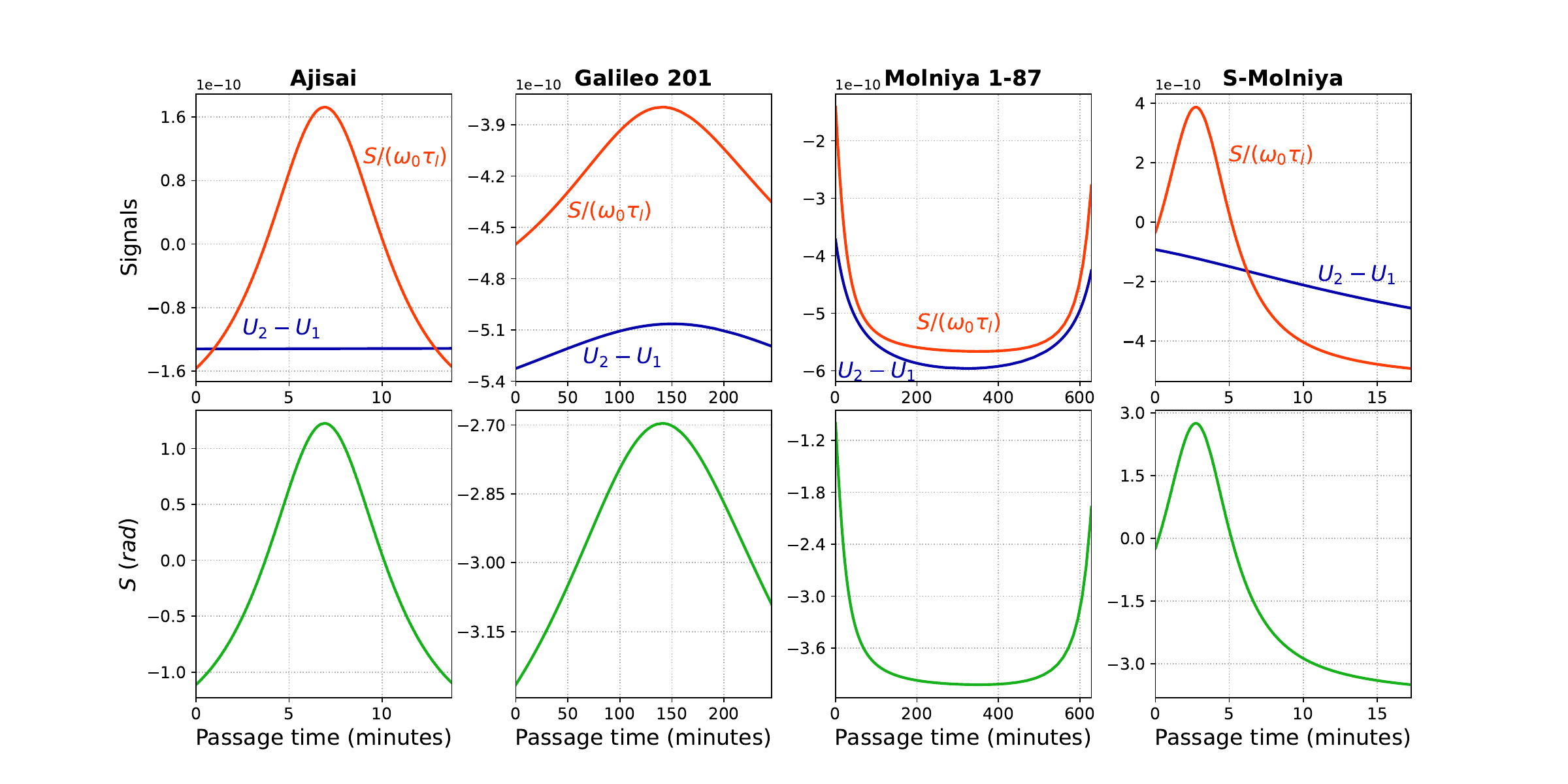}
  \caption{Results obtained with, from left to right, Ajisai (inclination 50$^\circ$, eccentricity 0.001, altitude 1490~km), Galileo 201 (inclination 50$^\circ$, eccentricity 0.158, altitude ranging from 17000 to 26210~km), Molniya 1-87 (inclination 63.6$^\circ$, eccentricity 0.68, altitude ranging from 2000 to 38000~km) and an hypothetical satellite on a South Pole Molniya-like orbit (the same parameters as Molniya 1-87, but with perigee on the northern hemisphere) seen from MLRO. Upper panels show the signals $\Delta U=U_2 - U_1$ and $S/(\omega_0 \tau_l)$ from Eq.~\eqref{FinalSignal_explicit}  (with $\alpha=0$) as a function of the passage time. Bottom panels show the signal $S$ expected with a delay line and wavelength {$\lambda = 1550$~nm.} }
\label{figure-simulation}
\end{figure*}

\section{Simulations}\label{s4}

We present the numerical estimation of the signal in Eq.~\eqref{FinalSignal_explicit}
 by exploiting the orbit of existing and simulated satellites, covering a wide range of orbital parameters.
 The first two satellites are currently used by the International Laser Ranging service (ILRS)~\cite{Pear2002}.
 The Satellite Laser Ranging (SLR) technique allows for a high accuracy estimation of the distance of such satellites by measuring the time-of-flight of laser pulses that are sent from a GS on Earth, then retroreflected by the CCRs mounted on the orbiting terminal, and finally collected by the same GS.
 ILRS makes available the Consolidated Prediction Format~\cite{CPF} files for SLR orbit, containing the geocentric (inertial Earth-centered) position of the satellites at a given time.
 We chose to perform the simulation using two satellites placed in different orbits: Ajisai (circular orbit) and Galileo 201 (eccentric orbit).
 In particular, Ajisai has an altitude of about 1500~km, as used in the estimation of the expected gravitational phase-shift after Eq.~\eqref{phigr}.
The used GS is the Matera Laser Ranging
Observatory (MLRO)~\cite{MLRO} of the Italian Space Agency, that was exploited for various demonstrations of the feasibility of satellite quantum communications~\cite{Vallone15,padova:15,padova:16,Vedovato17,Agnesi18,Calderaro18,Agnesi2019}.
 
 Two other simulations use satellites that are placed on a highly eccentric elliptical orbit, known as Molniya orbit.
 This orbit is well suited for telecommunications in polar regions and has therefore been exploited by the Soviet Union for placing its satellites.
 Satellites on these orbits spend most their time close to the apogee, with rapid passages at the perigee.
 We specialized our analysis on the Molniya 1-87 satellite~\cite{Molniya}, whose orbit has an inclination of 63.6$^\circ$ and an eccentricity of 0.68.
 Since all existing satellites placed on Molniya orbits are visible from the northern hemisphere only at perigee, we decided to simulate the orbit of a Molniya-like satellite spending most its time above the southern hemisphere and passing on top of the MLRO at the apogee (called S-Molniya).
 
All the orbits are simulated using the open source Orekit space dynamics library~\cite{orekit}, that can both simulate an orbit starting from the two-line elements (TLE) or the Keplerian orbital parameters and reproduce real passages as seen from an actual GS on Earth.

The upper panels of Fig.~\ref{figure-simulation} show the signal $S/(\omega_0\tau_l)$ from Eq.~\eqref{FinalSignal_explicit} as a function of the time passage for the satellites, while the bottom panel are the signals estimated by supposing that such terminals are equipped with an interferometer providing a delay line of {$l=1.2$~km (so $n=1.5$ implies $\tau_l \approx 6~\mu$s)} and that the initial wavelength is  {$\lambda = 2\pi c/\omega_0 = 1550$~nm}. This choice of the parameters $\tau_l$ and $\omega_0$  brings the strength of the signal in Eq.~\eqref{FinalSignal_explicit} into a measurable regime  on the order of few radians.

While the signal $S/(\omega_0\tau_l)$ is of the same order of magnitude for all the orbits, very low-eccentricity orbits for which $\Delta U = U_2 - U_1 \approx {\rm const}$ (e.g., Ajisai) are not suitable in practise, since the lack of variability in $\Delta U$ prevents its separation from the constant offset $\omega_0 \tau_l \delta_l$ that is due to the mismatch $\delta_l := (\tau_l^{\rm SC}-\tau_l^{\rm GS})/\tau_l^{\rm GS}$ of the delay lines, as discussed in  Appendix~\ref{appendix_dlines}.


\section{Conclusions}
Our proposal allows for the cancellation of the first-order Doppler effect in optical red-shift experiments. However, this proposal still faces two important practical issues. First, atmospheric turbulence is a limiting factor for large-distance optical interferometry. However, the planned temporal delay between the two pulses is four orders of magnitude lower than the conventional millisecond threshold of the turbulence correlation time~\cite{statop}. As a result, both the interfering beams suffer through the same random noise that is canceled in measuring $\varphi_{\rm SC}$  and $\varphi_{\rm GS}$. In fact, the same scale difference was successfully exploited in~\cite{padova:16}.

Second, the two delay lines cannot be perfectly identical. However, by exploiting commercially available fiber stretchers at each MZI and by monitoring in real-time the first-order interference with a stabilization laser of long coherence time (see Appendix~\ref{appendix_details} for more details), it is possible to  {phase-stabilize the two MZIs and}  achieve a   relative  precision  $\delta_l$ of the order of $10^{-6}$, which for {$l=1.2$~km} translates into an absolute difference of 1~mm.   {It is worth noticing that the capability of controlling with a  precision of 1~mm the relative length of two arms of 1~km of a balanced interferometer has been reported in~\cite{Xavier2011}, and this technique can be adapted to the case of unbalanced interferometers, provided an appropriate frequency reference to the two terminals (see Appendix~\ref{appendix_details}).} In this case the {measured} signal  gets a constant offset $\omega_0 \tau_l \delta_l$, that can be reliably estimated and eliminated by using SLR data.  Moreover, the additional variable term of the order $\delta_l$ can be eliminated similarly to the second-order Doppler terms (see Appendix~\ref{appendix_dlines}).

Concluding, in this work we propose an optical scheme to suppress the first-order Doppler effect in order to measure the gravitational red-shift with satellite systems. The possibility of testing gravitational physics using optical interferometric measurements between moving terminals represents an important point in the study of Einstein theory and it can open the way to new tests of its interplay with quantum mechanics through the exploitation of quantum optical effects. The recent advancements in satellite optical technologies make this proposal both attractive and feasible with current technologies.

\begin{acknowledgements}
The work of DRT is supported by the  grant   FA2386-17-1-4015 of AOARD. ARHS was supported by the Natural Sciences and Engineering Research Council of Canada and the Dartmouth College Society of Fellows. FV thanks Costantino Agnesi for useful discussions. We  acknowledge the International Laser Ranging Service (ILRS) for SLR data and software.
\end{acknowledgements}
\appendix

\section{More details on the experimental setup} \label{appendix_details}

{Here we provide some experimental details in order to attest the feasibility of our proposal. First, we address the problem of stabilizing two strongly unbalanced MZI guaranteeing, at the same time, that the two delay lines can be kept equal at the required precision. Second, since the MZIs have to be implemented with single-mode fibers to allow for strong imbalances and to achieve a good overlap of the interfering beams, we will sketch  a possible single-mode fiber-injection system exploiting adaptive optics. It is worth noticing that the proposed system is feasible with current technology given the maturity of fiber components at 1550~nm.}

{{\it Details of the interferometers} -- The MZI of both terminals employ two optical fibers (where one is a fiber spool much longer that the other) sandwiched between two 50/50 fiber beam splitters. In addition, one arm of the interferometer is equipped with a fiber stretcher (f.s.) in order to finely tune the imbalance to {$l=1.2$~km}. Note that a suitable laser emitting  pulses with short coherence time ($\approx 1$~ps) can be employed before the launch to ensure that the relative imbalance between the two delay lines is of the order of 1~mm, by measuring the  imbalance $\delta l$ of the single MZI~\cite{Vedovato17} with high-resolution superconducting nanowire single-photon detectors (SNSPDs). It is worth noticing that commercial fiber stretchers can provide down to $0.1$~$\mu$m of minimum step, so that, in principle,  $\delta l / l \approx 10^{-10}$. }

{To phase-stabilize the MZIs and keep the relative imbalances between the two to the required precision of $\delta_l = \Delta\tau_l / \tau_l \approx 10^{-6}$, an auxiliary stabilization (S) laser with central frequency $\nu_{\rm S}$ and  bandwidth $\Delta\nu_{\rm S}$ is employed at each terminal to monitor in real-time the first-order interference. 
The stabilization laser is assumed to be characterized by a coherence time $\tau_c^{\rm S}$ much longer than the target imbalance $\tau_l$, hence $\tau_c^{\rm S} \gg \tau_l$. Since $\tau_l = 6$~$\mu$s, a laser with a bandwidth of $\Delta  \nu_{\rm S} \ll 1 / \sqrt{4 \pi \tau_l^2} \approx 50$~kHz at a wavelength of, for example, 1560~nm is suitable for this task. With such a stabilization laser one can lock the optical phase  of the interferometer with a precision of the order of $\Delta \nu_{\rm S} / \nu_{\rm S} \approx 10^{-10}.$ 

Given the system described above, it is possible to ensure that the relative mismatch of the SC's delay line ($\tau_l^{\rm SC}$) with respect to the one of the GS ($\tau_l^{\rm GS} \equiv \tau_l$) is at most  $\delta_l = \Delta\tau_l / \tau_l \approx 10^{-6}$.  Having fixed $\tau_l = 6~\mu$s, we have that  $\Delta\tau_l \approx 10$~ps, and we can define the parameters of the signal source by requiring that 10~ps~$ < \tau_c \ll 1$~$\mu$s  to fulfill Eq.~\eqref{inequality_source}. Hence, a suitable signal source is a 1550~nm fiber-coupled laser with a repetition rate of 100 Hz, average power of 10~W (energy pulse of 100~mJ), coherence time $\tau_c$ of 10~ns, and linewidth of $\Delta \nu \approx 37.5$~MHz.
Recent experiments have demonstrated that such a source is feasible with current technology~\cite{Fix2011,Elsen2017}.
In order to reach the required optical precision $\delta \varphi_{gr} \approx 10~\mu$rad, it is necessary to detect a number of photons $N \gtrsim 1/\delta \varphi_{gr}^2 \approx 10^{10}$ photons.
Since a 100~mJ pulse at 1550~nm contains approximately $8 \cdot 10^{17}$ photons, the system can work with a level of losses up to almost 80~dB. Note that standard fibers at 1550~nm introduce a tolerable amount of losses even with strong imbalances, since the attenuation coefficient is about 0.2~dB/km at this wavelength. In order to achieve the required signal-to-noise ratio, it is necessary to use a low noise InGaAs photodiode.

{{\it Details of the fiber-injection system.---}The free-space propagation through the turbulent atmosphere affects the quality of the beam wavefront, which has to be corrected before being coupled to the SMF. To accomplish such a task we envisage to use an adaptive optics system like the one implemented in~\cite{Wright2015} and sketched in Fig.~\ref{figure-AO},  based on the exploitation of an additional beacon laser at a wavelength few nanometers apart from the signal one (e.g., 1545~nm). This additional beam share the same free-space optical path of the signal, and it is used as feedback for the adaptive optics system. Then, at the detection, it can be properly filtered out from the signal by using wavelength-division-multiplexer (WDM) filters, which provide down to 0.1~nm of bandwidth separation.}
\begin{figure}[!h]
  \includegraphics[width=0.4\textwidth]{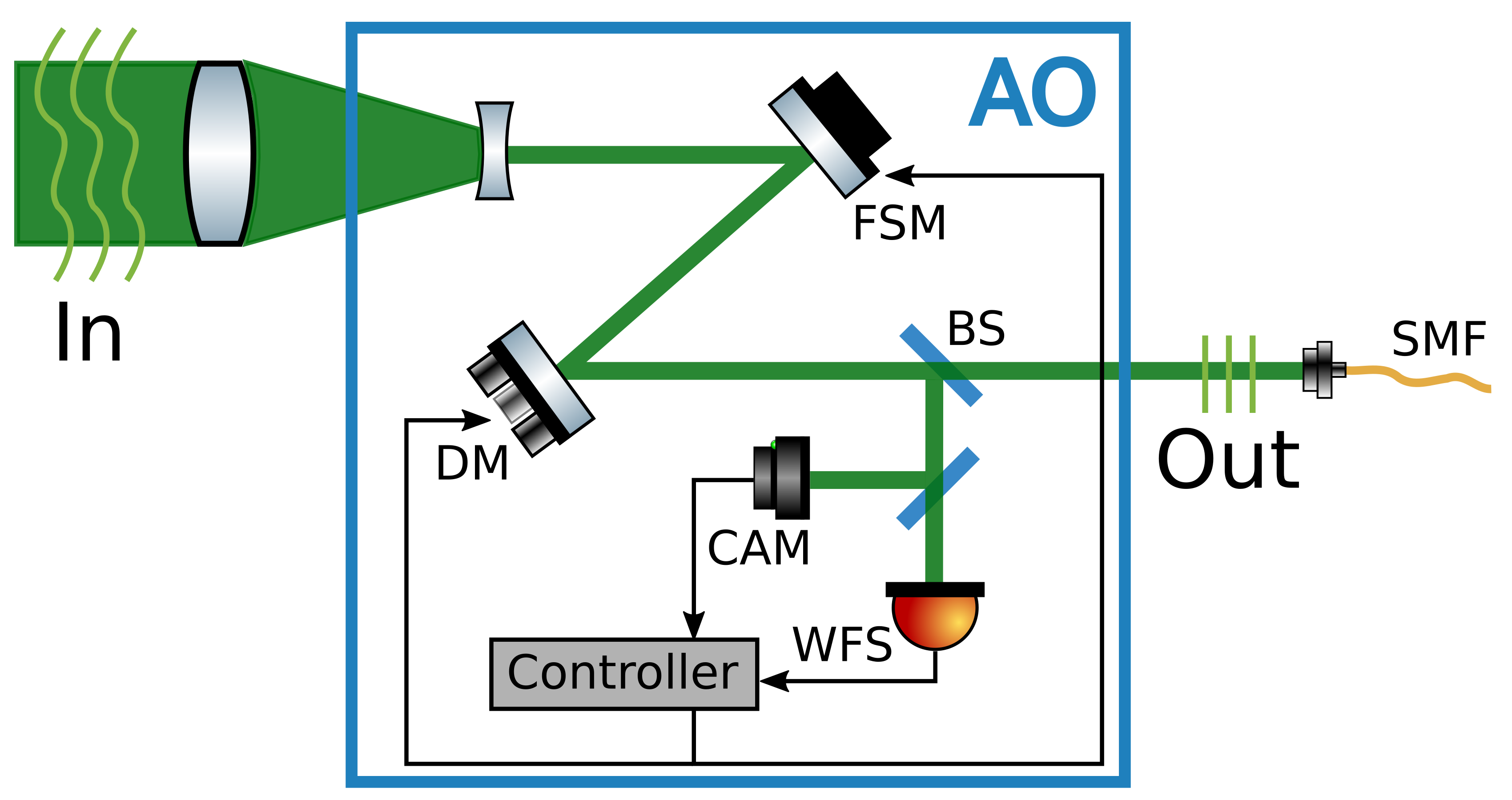}
  \caption{AO system needed for the free-space to SMF coupling.
  {The beam splitter (BM) can be replaced by a dichroic mirror if the beacon laser has a different wavelength with respect to the signal beam.}}
\label{figure-AO}
\end{figure} 
{Fig.~\ref{figure-AO} shows the expansion of the closed-loop adaptive optics (AO) box introduced in the top panel of Fig.~\ref{figure-scheme}. The input (In) of the AO system is the aberrated beam wavefront collected by a telescope (sketched as a lens), while the output (Out) is the corrected and collimated beam to be coupled to the SMF. The first element of the AO box is a lens whose focal length is chosen in order to reduce and collimate the incoming beam. The light is then reflected by a fast steering mirror (FSM) and a deformable mirror (DM) before passing a beam splitter (BS). The transmitted path exits from the AO box and provides the collimated and corrected beam to be coupled to the SMF, while the reflected path is collected by a camera (CAM) and a wavefront sensor (WFS). The CAM could be a camera or a position-sensitive-detector to measure the wandering of the beam at the focal plane and thus the low-order tilt due to turbulence, while the WFS could be a Shack-Hartmann sensor or a self-referenced interferometer to estimate the higher-order aberrations. The two signals generated by the CAM and the WFS drive the FSM and the DM in order to correct for low- and high-order aberrations of the wavefront.}

{The actual parameters of the AO box must be carefully chosen and they depend primarily on the level of expected turbulence, the dimensions of the beams, the optical power collected by the telescope and the velocity of the close-loop. In our scheme the working parameters of the two AO systems, one at the SC and the other at the GS, will be quite different, since the first has to correct the upgoing beam sent from the GS to the SC (about 40~dB of losses in a realistic scenario), while the other must be optimized for the go and return two-way path (about 80~dB of losses). However, since the AO system exploits an additional beacon with respect to the signal one, the required optical power is not an actual limitation for it to work.

As noticed in Ref.~\cite{Robert2016}, with long distance uplink propagation ($\gtrsim 1000$~km) the turbulence coherence area at the satellite receiver is much larger than the
typical receiver aperture size. In these cases, only a tip/tilt correction without AO on the satellite is sufficient for an optimal coupling into the single mode fiber.}

Since the optical payload of the SC and the required electronics comprise commercially available devices and telecom-compatible fiber technology, we can envisage that our proposal is feasible within a decade and with no prohibitive costs.

\section{Light propagation in the leading order PPN formalism and detailed calculation of the signal} 
\label{appendix_calculation}

\subsection{Resume and notation} \label{section_lightprop}

In the following we will use the convention $G = c = 1$ to simplify the notation.
An extended traitment of light propagation in the PPN formalism can be found in Ref.~\cite{will:93,wp:book}.  Light ray trajectories from $(t_\mrin, \vx_\mrin)$ to $(t, \vx$) (with PPN parameter $\gamma = 1$) are parametrized as
\be
\vx(t)=\vx_\mrin +\hn (t-t_\mrin)+\vx^{(2)}(t), 
\label{trajectory}
\ee
where $\vx^{(2)}(t)$ is the correction to the Newtonian straight propagation and the boundary condition gives $\vx^{(2)}(t_\mrin)=0$. Splitting $\vx^{(2)}(t)$ into its parallel  and perpendicular  component relative to $\hn$ as
\begin{align}
\vx^{(2)}_\| (t)&:= [\hn \cdot \vx^{(2)}(t)]\hn \equiv x^{(2)}_\|(t) \hn \ , \\ 
 \vx^{\,(2)}_\perp(t)&:=\vx^{(2)}(t) - \vx^{(2)}_\| (t)\ , 
\end{align}
then the two equations 
\begin{align}
\frac{dx^{(2)}_\|}{dt} &=-2 U , \\
\frac{d^2 \vx^{\,(2)}_\perp}{dt^2}&= 2\nabla U-2\hn(\hn\cdot\nabla U) , 
\end{align}
where the gravitational potential of a point-like Earth can be approximated by 
 \be
 U\approx U(r)\defeq\frac{M_{\oplus}}{|\vx_\mrin+\hn(t-t_\mrin)|}\equiv \frac{M_{\oplus}}{r},
 \ee
yield
\be
 \frac{d\vx^{\,(2)} }{dt}=-2 U(r)\hn-2\frac{M\vd}{d^{2}}  \left(\frac{\vx\cdot\hn}{r}-\frac{\vx_\mrin\cdot \hn}{r_\mrin}\right),
 \label{SeconOrderEOM}
\ee
where 
\be
\vd \defeq \hn \times ( \vx_\mrin \times \hn)= \vx_\mrin - (\hn\cdot\vx_\mrin)\hn,
\ee
is the vector joining the center of the Earth and the point of
closest approach of the unperturbed ray.   
Substituting Eq.~\eqref{trajectory} into Eq.~\eqref{SeconOrderEOM} and integrating from $t_\mrin$ to $t$ yields
\begin{align}
 \vx^{\,(2)} (t) &=  -2M\hn\ln\frac{(t-t_\mrin)+\hn\cdot\vx_\mrin+r(t)}{\hn\cdot\vx_\mrin+r_\mrin} \nonumber \\
&\quad -2    \frac{M\vd}{d^{2}} \left( r(t) - r_\mrin - \frac{\vx_\mrin\cdot \hn}{r_\mrin}(t-t_\mrin)\right).   \label{photon2}
\end{align}

\subsection{Light propagation for the one-way trips} \label{sec_back_oneway}
The set-up is depicted on Fig.~\ref{fig:mink}(center). With the precision of $\co(\epsilon^2)$, the trajectory of the SC is
\be
\vx_{\rm SC}(t) =  \vx_2+\vv_2(t-t_2)+\half\va_2(t-t_2)^2+\co(\epsilon^3) \ ,
\ee
hence by using Eq.~\eqref{eq_trasf_frame} we find
\be
\vx_{2^*}:= \vx_{\rm SC}(t_{2^*}) = \vx_2+\vv_2\tau_l+\half\va_2\tau_l^2+\co(\epsilon^3)\ .     \label{x2*}
\ee
Similarly the trajectory of the GS is
\be
\vx_{\rm GS}(t) =  \vx_1+\vv_1(t-t_1)+\half\va_1(t-t_1)^2+\co(\epsilon^3)\ ,
\ee
and so
\be
\vx_{1^*}:= \vx_{\rm GS}(t_{1^*})= \vx_1+\vv_1 t_{11^*}+\half\va_1 t_{11^*}^2+\co(\epsilon^3 ) \ . \label{x1*}
\ee
We comment on the relative importance of various terms at the end of this Section.

Given the parameters of the beam $A_2$ now we find the   (Euclidean) vector    $\hn_{1^*2^*}$, the new time-of-flight $T^*$, and  $t_{11^*}$ of the beam $A_1$ using their coincidence at $2^*$.
 For the emission from the GS we find the closest approach vector in Eq.~\eqref{photon2} is
 $\vd=\vx_{GS}$, with $\vx_{1}$ and $\vx_{1^*}$ for the respective pulses.
With the required precision we have
\be
\tau_l \equiv \tau_{22^*}^\mathrm{SC}=t_{22^*}\left(1-\half v_2^2-U_2\right) \ ,     \label{B13}
\ee
so that
\be
t_{22^*}=\tau_l\left(1+\half v_2^2+U_2\right), \label{B14}
\ee
and
\be
\tau_{11^*}^{\rm GS}=t_{11^*}\left(1-\half v_1^2-U_1\right) \ . \label{b15}
\ee
We expand the unknown quantities $T^*$, $t_{11^*}$ and $\hn_{1^*2^*}$ in powers of $\epsilon$ ($\epsilon\to 1$ at the end of the calculations)
\begin{align}
T^*&=T+\epsilon T_1+\epsilon^2 T_2 \ , \\
t_{11^*}&=\tau_l+\epsilon \delta_1+\epsilon^2\delta_2 , 
\end{align}
and
\be
\hn_{1^{\!*}2^{\!*}}=\hn_{12}+\epsilon\vnu^{\,*}_1+\epsilon^2\vnu^{\,*}_2 \ , 
\ee
which is required to have unit length, resulting in the conditions
\be
\hn_{12}\cdot\vnu^{\,*}_1=0, \qquad   2\hn_{12}\cdot\vnu^{\,*}_2+\nu_1^{*\,2}=0  \ .    \label{nuconst}
\ee

Expanding Eq.~\eqref{eq_time_oneway} in the quantities above in $\epsilon$ and equating terms of equal order yield
 \begin{align}
 \delta_1 &=- T_1 \ , \label{delt} \\ 
 \delta_2 &= - T_2 +  \tau_l\left(\half v_2^2+U_2\right) \label{deltas} \  .
 \end{align}

For the upward trajectory of the pulse  $\vx_\mrin=\vx_1$,  hence
\begin{align}
\vx_2&=\vx_1+ \hn_{12}T + \vec\chi_\uparrow(\vx_1,\hn_{12},T) \ ,      \label{x2}
\end{align}
where we re-wrote the $\co(\epsilon^2)$ terms coming from Eq.~\eqref{photon2} as
\begin{align}
\vec\chi_\uparrow(\vx_1,\hn_{12},T)&\defeq -2M\hn_{12}\ln\frac{T+\hn_{12}\cdot\vx_1+|\vx_1+ \hn_{12}T|}{\hn_{12}\cdot\vx_1+r_1} \nonumber\\
&-2    \frac{M\vd_1}{d_1^{2}} \left( |\vx_1+ \hn_{12}T| - r_1- \frac{\hn_{12}\cdot\vx_1 }{r_1}T\right),
\end{align}
{with $\vec{d}_1=\hn_{12}\times(\vx_1\times \hn_{12})$.}

The delayed pulse $A_1$ leaves the GS at $1^*$. Using the parameters specifying its trajectory and noting that $r_1\equiv r_{1^*}$, as well as that the post-Newtonian corrections to the light trajectory are already of the order of $\epsilon^2$ --- so that the corrections due to difference in $\hn_{1^{\!*}2^{\!*}}$ and $T^*$ from  $\hn_{12}$ and $T$, respectively, are of the order $\epsilon^3$ and can be ignored ---  it follows that 
\begin{align}
 \vx_{2^*}&=\vx_{1^*}+ \hn_{1^{\!*}2^{\!*}}T^*+ \vch_\uparrow(\vx_1,\hn_{12},T) \nonumber \\
 &= \vx_1+\vv_1 t_{11^*}+\half\va_1 t_{11^*}^2 + \hn_{1^{\!*}2^{\!*}}T^*+ \vch_\uparrow(\vx_1,\hn_{12},T) \label{x2*2}
\end{align}

In order to find the expressions for the unknown quantities $T^*$, $t_{11^*}$ and $\hn_{1^*2^*}$ up to the second order, we match Eq.~\eqref{x2*} --- using $\vx_2$ given by Eq.~\eqref{x2} --- with the above Eq.~\eqref{x2*2}.
As a result
\be
\hn_{12}T+\vv_2\tau_l+\half\va_2\tau_l^2=\vv_1t_{11^*}+\half\va_1 t_{11^*}^2 +\hn_{1^{\!*}2^{\!*}} T^*.
\ee
Expanding it order-by order in $\epsilon$ results in the final six equations
\be
 \vv_2\tau_l= \vv_1 \tau_l+\hn_{12}T_1 + \vnu^{\,*}_1 T,      \label{trip1}
\ee
 and
 \be
\half\va_2\tau_l^2=-\vv_1 T_1+\half \va_1\tau_l^2+\vnu^{\,*}_2 T+\vnu^{\,*}_1 T_1+\hn_{12}T_2,   \label{trip2}
 \ee
   where we used   Eq.~\eqref{delt}.    Using the first of the relations \eqref{nuconst} with \eqref{trip1} results in
\be
T_1=\hn_{12}\cdot(\vv_2-\vv_1)\tau_l= (\md_2-\md_1)\tau_l,
\ee
and
\be
\vnu^{\,*}_1=(\vv_2-\vv_1)\frac{\tau_l}{T}-\hn_{12}\frac{T_1}{T}=  \frac{\tau_l}{T}\left((\vv_2-\vv_1)-\hn_{12} (\md_2-\md_1)\right), \ee
{ where $\md_k \defeq \hn_{12}\cdot\vv_k$.}
We see that $T_1/T\sim|\vnu^{\,*}_1|=\co(\epsilon\mu)$.

 Using the second of the relations \eqref{nuconst} the second triple of the equations \eqref{trip2} results in
 \begin{align}
  T_2 &=\half\hn_{12}\cdot(\va_2-\va_1)\tau_l^2+\md_1 T_1+\half\nu_1^{*2} T \nonumber \\ &=
   \half (\ma_2-\ma_1)\tau_l^2 +\md_1(\md_2-\md_1)\tau_l \nonumber \\&\quad+ \frac{\tau_l^2}{2T}\left((\vv_2-\vv_1)^2-(\md_2-\md_1)^2\right), \label{t2}
 \end{align}
 where $\mathfrak{a}_k\defeq \hn_{12}\cdot\va_k$,
and
\be
\vnu^{\,*}_2=\big(\half\va_2\tau_l^2+\vv_1 T_1-\half \va_1\tau_l^2- \vnu^{\,*}_1 T_1-\hn_{12}T_2\big)/T.
\ee

Note that in our setting the second term on the right-hand-side of Eq.~\eqref{t2} dominates the other two by the factor of the order $T/\tau_l = \mu^{-1} \approx  {10^3}$. Even so, the sub-dominant terms are an order of magnitude larger than $\epsilon^3$, and hence should be kept. The terms proportional to $\tau_l^2$ are absent from the expressions in Ref.~\cite{phase-freq} where it was assumed that $\tau_l\lesssim\epsilon$.

Using Eqs.\eqref{delt}-\eqref{deltas} we get
\begin{align}
t_{11^*}&=  \tau_l + \delta_1 + \delta_2 \nonumber\\
&= \tau_l - T_1 - T_2 + \tau_l(\half v_2^2 + U_2) \nonumber \\
&=\tau_l\big(1+\half v_2^2+U_2\big)-T_1-T_2 \ ,
\label{eq:tau_11star}
\end{align}
which is related to $\tau_{11^*}^{\rm GS}$ by Eq.~\eqref{b15}. Hence
\be
\tau_{11^*}^\mathrm{GS}-\tau_l=-T_1+\tau_l\left(\half(v_2^2- v_1^2)+U_2-U_1\right)-T_2,
   \ee
where the term $T_1$ is responsible for the first-order Doppler effect in the phase-difference at the SC
   \be
  \varphi_\mathrm{SC}=  \omega_0(\tau_{11^*}^\mathrm{GS} - \tau_l) =  -\omega_0 T_1+ \varphi_{\rm SC}^{(2)}
\ee
with  
\be
 \varphi_{\rm SC}^{(2)} = \omega_0\tau_l\left(\half(v_2^2- v_1^2)+U_2-U_1-T_2/\tau_l\right).
   \ee

\subsection{Light propagation for the two-way trips} \label{sec_back_twoway}
{\it Definition of the relevant quantities.---}The set-up is depicted on Fig.~\ref{fig:mink}{(right)}. The initial parameters of the beam $B_2$ are known, and we use them to express parameters of the beam $B_1$. The relevant parameters for the $2\to 3$ part of the trajectory are
   the propagation direction
 \be
 \hn_{23}=-\hn_{12}+\epsilon \vnu_1+\epsilon^2\vnu_2 \ ,
 \ee
 that satisfies the relations
\be
\hn_{12}\cdot\vnu_1=0\ , \qquad   2\hn_{12}\cdot\vnu_2+\nu_1^{2}=0\ ,
\ee
and the time-of-flight from the SC to the GS
\be
 P=T+\epsilon\Delta_1+\epsilon^2\Delta_2 \ .
 \ee
 
 Our primary object of interest is the time of departure of the pulse $B_1$ (in general it departs at the moment $t_{\bar 1}\neq t_{1^*}$).
This time can be decomposed as
\be
t_{1\bar{1}} =\tau_l+\epsilon\bar\delta_1+\epsilon^2\bar \delta_2 \ ,
\ee
and the position of the GS at the moment $t_{\bar{1}}$ is
\be
\vx_{\bar 1 }= \vx_1+\vv_1 t_{1\bar{1}}+\half\va_1 t_{1\bar{1}}^2 \ .
\ee
The quantity $t_{1\bar{1}}$ will be recovered from the coincidence of the beams $B_1$ and $B_2$ at $3^*$.

The flight time from $t_{\bar{1}}$ to $t_{\bar{2}}$ is
\be
 \bar T=T+ \epsilon \bar T_1+\epsilon^2\bar T_2 \ ,
 \ee
and   the launch direction is given by
\be
\hn_{\bar 1\bar 2}=\hn_{12}+\epsilon \underbar{$\vnu$}_{\,1}+\epsilon^2\underbar{$\vnu$}_{\,2}\ .
\ee
The pulse reflected at $t_{\bar{2}}$ is directed along
 \be
\hn_{\bar 23^*}=-\hn_{12}+\epsilon{\vnu\,}'_1+\epsilon^2{\vnu\,}'_2\ ,
\ee
and the travel takes 
\be
P'=T+\epsilon \Delta_1'+\epsilon^2\Delta_2' \  .
\ee

{\it $2\to3$ parameters.---}Six independent parameters are obtained from the expressions for $\vx_3$. On the one hand,  the GS motion implies
 \begin{align}
 \vx_3&=\vx_1+\vv_1(T+P)+\half \va_1(T+P)^2\nonumber\\&=\vx_1+2\vv_1 T+\vv_1\Delta_1+2\va_1 T^2 \ .
 \end{align}

Noting for the downward motion the closest distance to the centre of the Earth is still $r_3\equiv r_1$,
the correction to the trajectory is
\begin{align}
\vch_\downarrow &:= \vch_\downarrow(\vx_2,\hn_{23},P)=2M\hn_{12}\ln\frac{r_1-\hn_{12}\cdot\vx_1}{r_2-\hn_{12}\cdot\vx_2} \nonumber\\
&-2    \frac{M {\vd_1}}{{d_1^2}} \left(r_1 - r_2 + \frac{\hn_{12}\cdot\vx_2 }{r_2}T\right)
\end{align}
We also rewrite
\begin{align}
\vch_\uparrow&:=\vch_\uparrow(\vx_1,\hn_{12},T)\defeq -2M\hn_{12}\ln\frac{r_2+\hn_{12}\cdot\vx_2}{r_1+\hn_{12}\cdot\vx_1}
\nonumber\\&-2    \frac{M{\vd_1}}{d_1^{2}} \left( r_2- r_1 - \frac{\hn_{12}\cdot\vx_1 }{r_1}T\right).
\end{align}
In the above expressions we use $\vx_2=\vx_1+\hn_{12}T$.

As a result the expression for $\vx_3$ that is obtained by following the light pulse is
\begin{align*}
\vx_3&=\vx_1+\hn_{12}T+\hn_{23}P+\vch_\uparrow+\vch_\downarrow\nonumber\\&=\vx_1-\hn_{12}\Delta_1+\vnu_1 T \nonumber\\&\quad-\hn_{12}\Delta_2+\vnu_1\Delta_1+\vnu_2 T+\vch_\uparrow+\vch_\downarrow.
\end{align*}
  The first-order terms
 \be
 2\vv_1 T=\vnu_1 T-\hn_{12}\Delta_1
 \ee
 lead to
 \be
 \Delta_1=-2\md_1 T,
 \ee
 and
 \be
 \vnu_1=-2\hn_{12}\md_1+2\vv_1.
 \ee
 The second order equation is
 \be
 \vv_1\Delta_1+2\va_1 T^2=-\hn_{12}\Delta_2+\vnu_1\Delta_1+\vnu_2 T +\vch_\uparrow+\vch_\downarrow
 \ee
 that results in
 \be
 \Delta_2=-\md_1\Delta_1-2\ma_1 T^2-\half\nu_1^2 T+\chi_\uparrow+\chi_\downarrow,
 \ee
where $\chi_{\uparrow,\downarrow}\defeq\hn_{12}\cdot\vch_{\uparrow,\downarrow}$ and
\be
\vnu_2 = (\vv_1\Delta_1+2\va_1 T^2+\hn_{12}\Delta_2-\vnu_1\Delta_1-\vch_\uparrow-\vch_\downarrow)/T.
\ee

{\it $\bar 1\to\bar 2$ parameters.---}At the order $\epsilon^2$ the two expressions for the SC position $\vx_{\bar 2}$ are
 \begin{align}
\vx_{\bar 2}&=\vx_{\bar 1}+\hn_{\bar 1\bar 2}\bar T+\vec\chi_\uparrow(\vx_{\bar 1},\hn_{\bar 1 \bar 2},\bar T)\nonumber\\&=
\vx_{\bar 1}+\hn_{12}T+\underbar{$\vnu$}_{\,1}T+\hn_{12}\bar T_1\nonumber\\&\quad+\underbar{$\vnu$}_{\,2} T+\underbar{$\vnu$}_{\,1}\bar T_1+\hn_{12}\bar T_2+\vec\chi_\uparrow(\vx_1,\hn_{12},T),
 \end{align}
where
\be
\vx_{\bar 1 }= \vx_1+\vv_1(\tau_l+\bar\delta_{ 1})+\half\va_1\tau_l^2.
\ee
and  
\begin{align}
 \vx_{\bar 2}&=\vx_2+\vv_2(t_{\bar 2}-t_2)+\half \va_2(t_{\bar 2}-t_2)^2\nonumber\\&=\vx_1+ \hn_{12}T \nonumber\\&\quad+ \vec\chi_\uparrow(\vx_1,\hn_{12},T)+\vv_2(\tau_l+\bar\delta_1+\bar T_1)+\half \va_2\tau_l^2,   \label{x2b}
 \end{align}
as this is where the SC is at the moment $t_{\bar 2}=t_{\bar 1}+\bar T$.   The first six equations (that contain seven variables) are
\be
 \vv_1\tau_l+\underbar{$\vnu$}_{\,1}T+\hn_{12}\bar T_1=\vv_2\tau_l
\ee
at the order of $\epsilon$ and
\begin{align}
 \vv_1\bar\delta_{1}+\half\va_1\tau_l^2+\underbar{$\vnu$}_{\,2} T+&\underbar{$\vnu$}_{\,1}\bar T_1+\hn_{12}\bar T_2=\nonumber\\ &=\vv_2(\bar\delta_1+\bar T_1)+\half \va_2\tau_l^2 \label{bt2e}
\end{align}
 at the order $\epsilon^2$. We get from the first-order equations (that are self-contained)
 \begin{align}
 \bar T_1&=(\md_2-\md_1)\tau_l\equiv T_1 \ , \\   \underbar{$\vnu$}_{\,1}&= \big((\vv_2-\vv_1)-\hn_{12}(\md_2-\md_1)\big)\tau_l/T\equiv \vnu_1^{\,*}\ .
 \end{align}
 
 {\it $\bar 1\to \bar 2\to 3^*$ vs $1\to 2\to 3\to 3^*$.---}Since the difference between $\tau_l \equiv \tau_{33^*}^{\rm GS}$ and $t_{33^*}$ is of the order of $\epsilon^2$, we have 
 \be
 \vx_{3^*}=\vx_3+\vv_3 \tau_l+\half\va_3\tau_l^2 \ ,
 \ee
 where the 2nd order expression for $\vv_3$ is
 \be
 \vv_3=\vv_1+2\va_1T.
 \ee
Since $\tau_l/T\sim \mu \sim \epsilon/10$ we discard the term in the correction of the velocity and since $\va_3=\va_1+\co(\epsilon^3)$ we just set
$\va_3=\va_1$.
Hence
 \begin{align}
\vx_{3^*}&=\vx_1-\hn_{12}\Delta_1+\vnu_1 T -\hn_{12}\Delta_2+\vnu_1\Delta_1+\vnu_2 T\nonumber\\&\quad+\vec\chi_\uparrow+\vec\chi_\downarrow+\vv_1\tau_l+\va_1\tau_l(2T+\half \tau_l). 
\end{align}
It should be matched with
\begin{align}
\vx_{3^*}&=\vx_{\bar 2}+\hn_{\bar 23^*}P'+\vec\chi_\downarrow(\vx_{\bar 2},\hn_{\bar 23^*},P')\nonumber\\&=
\vx_{\bar 2}-\hn_{12}T-\hn_{12}\Delta_1'+{\vnu\,}'_1T\nonumber\\&\quad-\hn_{12}\Delta_2'+{\vnu\,}'_1\Delta_1'+{\vnu\,}'_2T  +\vec\chi_\downarrow(\vx_{2},\hn_{23},P) \ , 
\end{align}
that, by using  Eq.~\eqref{x2b}, becomes
\begin{align}
\vx_{3^*}&=\vx_1 + \vec\chi_\uparrow+\vv_2(\tau_l+\bar\delta_1+\bar T_1)+\half \va_2\tau_l^2 \nonumber\\&\quad-\hn_{12}\Delta_1'+{\vnu\,}'_1T-\hn_{12}\Delta_2'+{\vnu\,}'_1\Delta_1'+{\vnu\,}'_2T  +\vec\chi_\downarrow \ .
\end{align}

From the coincidence of the positing $\vx_{3*}$ we obtain further six equations,
\be
-\hn_{12}\Delta_1+\vnu_1 T +\vv_1\tau_l=   -\hn_{12}\Delta_1'+{\vnu\,}'_1T  +\vv_2\tau_l,    \label{x3*1} \ee
  and
  \begin{align}
   -&\hn_{12}\Delta_2+\vnu_1\Delta_1+\vnu_2 T+\va_1\tau_l(2T+\half \tau_l) = \nonumber\\ &=   -\hn_{12}\Delta_2'+{\vnu\,}'_1\Delta_1'+{\vnu\,}'_2T +
   \vv_2(\bar\delta_1+\bar T_1)+\half \va_2\tau_l^2. \label{x3*2}
   \end{align}

We note that from Eq.~\eqref{eq_trasf_frame} we have
\begin{equation}
\tau_l \equiv \tau_{33^*}^{\rm GS} = t_{33^*}\left(1-\half v_1^2-U_1\right)
\end{equation}
so that
\be
t_{33^*}=\tau_l\left(1+\half v_1^2+U_1\right)
\ee
and the final equations obtained by using the $\epsilon$-expanded quantities in Eq.~\eqref{eq_time_twoway} are
\be
\Delta_1=\bar T_1+\Delta_1'+\bar\delta_1 \ee
and
\be
\Delta_2+\tau_l\left(\half v_1^2+U_1\right)=\bar T_2+\Delta_2' +\bar\delta_2  \ . \label{bdo2}
\ee
From Eq.~\eqref{x3*1} we get
\be
\Delta_1'     =\Delta_1+(\md_2-\md_1)\tau_l  =-2\md_1 T       +(\md_2-\md_1)\tau_l
\ee
(note that $(\Delta'-\Delta)/T\sim \co(\epsilon \mu)$) and
\begin{align}
\vnu_1'&=\vnu_1+(\hn_{12}(\Delta_1'-\Delta_1)  +(\vv_1-\vv_2)\tau_l)/T\nonumber\\&=  \vnu_1+(\hn_{12}(\md_1-\md_2)+ \vv_1-\vv_2  )\tau_l/T\nonumber\\&=\vnu_1-\vnu_1^{\,*} \ , \label{nu1p}
\end{align}
and calculation of the first-order terms is completed by
\be
\bar\delta_1=\Delta_1-\Delta_1'-\bar T_1=-2(\md_2-\md_1)\tau_l\equiv 2\delta_1=-2T_1.
\ee
This is the basis for the Doppler cancellation scheme.

Now we can use Eq.~\eqref{bt2e} to obtain
\begin{align}
\bar T_2&=-2\md_1\delta_{ 1}-\half\ma_1\tau_l^2+\half\underbar{$\nu$}_{\,1}^2 T+\md_2 \delta_1 +\half \ma_2\tau_l^2 \nonumber\\
&=(\md_2-\md_1)(2\md_1-\md_2)\tau_l+\half(\ma_2-\ma_1)\tau_l^2\nonumber\\&\quad+\half\big((\vv_2-\vv_1)^2-(\md_2-\md_1)^2)\tau_l^2/T.
\end{align}

From Eq.~\eqref{x3*2} we get
\begin{align}
-\Delta_2-&\half\nu_1^2 T+\ma_1\tau_l(2T+\half \tau_l)  = \nonumber\\&=   -\Delta_2'-\half\nu_1'{}^{2}T +
   \md_2(2\delta_1+ T_1)+\half \ma_2\tau_l^2.
\end{align}
We further use $2\delta_1+T_1=\delta_1=-(\md_2-\md_1)\tau_l$ and using Eq.~\eqref{nu1p} we obtain
\be
\nu_1'{}^{2}=\nu_1^2+2\vnu_1\cdot(\vv_1-\vv_2)\tau_l/T+\co(\epsilon^2 \mu^2) \ , 
\ee
leading to
\be
\Delta_2'=-2\ma_1\tau_lT+  \Delta_2-\vnu_1\cdot(\vv_1-\vv_2)\tau_l +
   \md_2 \delta_1 +\half (\ma_2-\ma_1)\tau_l^2 \ , 
\ee
that reduces to
\begin{align}
\Delta_2-\Delta_2' &= 2\ma_1\tau_lT+  \vnu_1\cdot(\vv_1-\vv_2)\tau_l  \nonumber\\ &\quad+\md_2 (\md_2-\md_1)\tau_l -\half (\ma_2-\ma_1)\tau_l^2 \nonumber\\
   &= 2\ma_1\tau_lT+ (2\md_1+\md_2)(\md_2-\md_1)\tau_l\nonumber\\&\quad+2\vv_1\cdot(\vv_1-\vv_2)\tau_l -\half (\ma_2-\ma_1)\tau_l^2\ .
\end{align}

{\it Phase-difference at the GS.---}By noting that
\begin{align}
\tau_{1\bar{1}}^\mathrm{GS} &= t_{1\bar{1}}(1 - \half v_1^2 - U_1) \nonumber \\
&= (\tau_l - 2 T_1 + \bar{\delta}_2)(1 - \half v_1^2 - U_1)\nonumber\\ &=\tau_l - 2 T_1 +\big(\bar \delta_2- \tau_l\left(\half v_1^2+U_1\right)\big) \ ,
\label{eq:tau11bar}
\end{align}
we can write
\begin{align}
 \tau_{1\bar{1}}^\mathrm{GS} - \tau_l  
&= -2 T_1+\big(\bar \delta_2- \tau_l\left(\half v_1^2+U_1\right)\big)\nonumber\\ &\eqdef -2T_1+ \Delta^{(2)}\ , 
\end{align}
where $\Delta^{(2)}$ is defined according to Eq.~\eqref{bdo2} as
\begin{align}
\Delta^{(2)}&:=\Delta_2-\Delta'_2-\bar T_2\nonumber \\&=2\ma_1\tau_lT+ 2\md_2(\md_2-\md_1)\tau_l+2\vv_1\cdot(\vv_1-\vv_2)\tau_l \nonumber\\&- (\ma_2-\ma_1)\tau_l^2
-\half\big((\vv_2-\vv_1)^2-(\md_2-\md_1)^2)\tau_l^2/T.
\end{align}

In the end, the phase-difference at the GS results
 \be
  \varphi_\mathrm{GS}=  \omega_0 (\tau_{1\bar{1}}^\mathrm{GS} - \tau_l)=  -2\omega_0 T_1+ \varphi_\mathrm{GS}^{(2)}
\ee
with
\be  
  \varphi_\mathrm{GS}^{(2)} = \omega_0\Delta^{(2)}\ .
   \ee
{\it The signal.---}Having the explicit expressions for $\varphi_{\rm SC}$ and $\varphi_{\rm GS}$ up to the second order, we obtain
\begin{align}
 S&= \varphi_\mathrm{SC}-\half \varphi_\mathrm{GS} =\varphi_\mathrm{SC}^{(2)}-\half \varphi_\mathrm{GS}^{(2)}\nonumber\\
 &=\omega_0\left(\tau_l[\half(v_2^2- v_1^2)+U_2-U_1]-T_2-\half\Delta^{(2)}\right)
   \end{align}
that explicitly reads
\begin{align}
\frac{S}{\omega_0\tau_l}&=U_2-U_1+\half(v_2^2- v_1^2)\nonumber\\
&\quad-\vv_1\cdot(\vv_1-\vv_2) - (\md_2^2-\md_1^2)-\ma_1T \nonumber\\&\quad\quad- \frac{\tau_l}{4T}\left((\vv_2-\vv_1)^2-(\md_2-\md_1)^2\right)
\end{align}
and leads to Eq.~\eqref{FinalSignal_explicit}.

\section{Unequal delay lines} \label{appendix_dlines}

As discussed above, it is impossible for two delay lines to be perfectly identical. We characterize {the} difference in the proper propagation times as
\be
\tau_l^\mathrm{GS}=\tau_l, \qquad \tau_l^\mathrm{SC}=\tau_l+ {\Delta  \tau_l} := \tau_l (1 + \delta_l) \ ,
\ee
and we assume that the relative difference of the delay lines is at most of the order
\begin{equation}
 \delta_l = \frac{\tau_l^{\rm SC} - \tau_l^{\rm GS}}{\tau_l^{\rm GS}}=\frac{{\Delta  \tau_l}}{\tau_l} \lesssim  {10^{-6} \sim \epsilon^{6/5}} \ .
\end{equation}

The analysis of the two-way trip (Sec.~\ref{sec_back_twoway}) does not change. On the other hand, for the one-way trip (Sec.~\ref{sec_back_oneway}) we now have instead of Eq.~\eqref{B13} the following relation
\be
\tau_l(1 + \delta_l) \equiv \tau_{22^*}^\mathrm{SC}=t_{22^*}\left(1-\half v_2^2-U_2\right) \  ,
\ee
so that Eq.~\eqref{B14} becomes
\be
t_{22^*}=\tau_l  \left(1+\delta_l+\half v_2^2+U_2\right) \ .
\ee
The rest of the calculations proceed as before, resulting in the departure coordinate time (GRF) of the beam $A_1$.
Using the results for $T_1$ and $T_2$ we get
\be
t_{11^*}=\tau_l\big(1+\delta_l + \half v_2^2+U_2\big)-T_1-T_2.
\ee
Accordingly, 
\begin{align}
\ \tau_{11^*}^\mathrm{SC}-\tau_l =-T_1+\tau_l\left(\delta_l+\half(v_2^2- v_1^2)+U_2-U_1\right)-T_2,
   \end{align}
where the term $T_1$ is responsible for the first-order Doppler effect in the phase-difference at the SC
\begin{equation}
 {
    \varphi_\mathrm{SC}= \omega_0(\tau_{11^*}^\mathrm{SC} - \tau_l)=   -\omega_0 T_1 + \omega_0 \tau_l \delta_l + \varphi_{\rm SC}^{(2)}} \ ,
\end{equation}
$\omega_0 \tau_l \delta_l$ a constant offset, and the higher order corrections related to the mismatch of the delay times are at least of the order $\co(\epsilon^3)$.

Effects of the constant delay line mismatch $\Delta l=l\delta_l$ can be removed by the data processing. However, a random time-varying mismatch can wash out the imprints of the gravitational redshift (and the second-order effects in general). The most immediate source of randomness are the temperature fluctuations that give
\be
\Delta\delta_l\approx\kappa\Delta\mathbbm{T},
\ee
where $\kappa$ is the thermal expansion coefficient and $\Delta\mathbbm{T}$ is the onboard temperature fluctuation during one passage of the satellite. Given the results of Section.~\ref{s4} having $\Delta\delta_l\sim 10^{-11}$ allows identification of the second order effects, and $\Delta\delta_l\sim 10^{-13}-10^{-15}$ the precision measurements of the gravitational red shift.  If $\kappa\sim 10^{-7}-10^{-9} \, \mathrm{K}^{-1}$
and the maximal temperature variation $\Delta\mathbbm{T}\sim 10^{-5}$~K as in the desiderata list of the ORTIS mission \cite{ORTIS:01}, then not only identification of $\Delta U$, but also putting the EM-based bounds on $\alpha$ is possible. Current results from the pathfinder missions \cite{LISA:19,Taiji:21} reliably set $\Delta\mathbbm{T}\lesssim 10^{-3}$, bringing the all-optical measurement of the gravitational red shift into the realm of possibility.

\section{Interference visibility} \label{appendix_visibility}
Besides the spatial overlap of the interfering beams (that is granted by the use of single mode fibers at the two terminals), the different arrival time at the detector can cause a decrease in the interferometric visibility $\mathcal{V}$, limiting the precision of phase estimation to $\delta \varphi_{gr} \approx 1/(\mathcal{V}\sqrt{N})$, where $N$ is the number of detected photons.
Since we are dealing with optical pulses whose linewidth is much smaller than the central frequency, it is possible to perform all the calculations in the monochromatic approximation used in Section~\ref{PPN-1} and evaluate the visibility by looking at the overlap between the back-propagated pulses at the two different starting points (in the GS reference frame).

Following the conventions of~\cite{padova:16}, we define the envelope function of a Gaussian pulse centered in $t_A$ as
\begin{equation}
    \mathcal{A}_{t_A}(t) = \sqrt[4]{\frac{1}{\pi \tau_c^2}} \exp\left[-\frac{(t-t_A)^2}{2\tau_c^2}\right],
\end{equation}
where $\tau_c$ is the coherence time of the pulse.
The overlap between two pulses centered, respectively, in $t_A$ and $t_B$, is given by the integral
\begin{align}
    \mathcal{V} &= \sqrt{ \frac{1}{\pi \tau_c^2} } \int dt \exp\left[-\frac{(t-t_A)^2}{2\tau_c^2}\right] \exp\left[-\frac{(t-t_B)^2}{2\tau_c^2}\right] \nonumber \\
    &= \exp\left[-\frac{(t_A-t_B)^2}{4\tau_c^2}\right].
\end{align}

From this formula and using the conventions of Section~\ref{PPN-1}, it is possible to calculate the visibility in the one-way and in the two-way configuration, as
\begin{align}
    \mathcal{V}_{\rm one-way} &= \exp\left[ -\frac{\left(\tau^{\rm GS}_{11^{*}} - \tau_l\right)^2}{4\tau_c^2} \right] \ , \\ 
    \mathcal{V}_{\rm two-way} &= \exp\left[ -\frac{\left(\tau^{\rm GS}_{1\bar{1}} - \tau_l\right)^2}{4\tau_c^2} \right] \ .
\end{align}
By inserting, respectively, Eq.~(\ref{eq:tau_11star}) and Eq.~(\ref{eq:tau11bar}) in the above equations, we obtain at the leading order
\begin{align}
    \mathcal{V}_{\rm one-way} &\approx \exp{\left[ -\frac{T_1^2}{4\tau_c^2} \right]} = \exp{\left[ -\frac{(\md_2-\md_1)^2 \tau_l^2}{4\tau_c^2} \right]} \ , \\ 
    \mathcal{V}_{\rm two-way} &\approx \exp{\left[ -\frac{T_1^2}{\tau_c^2} \right]} = \exp{\left[ -\frac{(\md_2-\md_1)^2 \tau_l^2}{\tau_c^2} \right]} \ . 
\end{align}
As evident from the previous equations, the visibility depends on $( \tau_l / \tau_c)^2$.
We have verified that, given an imbalance $\tau_l~=~6~\mu$s and a coherence time of $\tau_c~=~10$~ns, the visibility is higher than $99\%$ for all studied trajectories and its effect can be neglected.

\end{document}